\documentclass[journal,a4paper,onecolumn,11pt]{IEEEtran} 
\usepackage{amssymb,amsfonts,amstext}
\usepackage{verbatim}
\usepackage{amsmath}
\usepackage{bm}
\usepackage{color}
\usepackage{algorithm}
\usepackage{algorithmic}
\usepackage{latexsym}
\usepackage{multirow}

\DeclareMathOperator{\tr}{tr}

\usepackage[dvips]{graphicx}

\newtheorem{proposition}{Proposition} 
\newtheorem{theorem}{Theorem}

\newtheorem{definition}{Definition}

\newcommand{\ZZ}{\mathbb{Z}}
\newcommand{\bA}{\mathbb{A}}

\newcommand{\FF}{\mathbb{F}}

\newcommand{\Tr}{\mathrm{Tr}}

\newcommand{\be}{\begin{equation}}
\newcommand{\ee}{\end{equation}}

\def\sX{\mathsf{X}}
\def\sZ{\mathsf{Z}}
\def\sA{\mathsf{A}}

\newenvironment{proofof}[1]{\vspace*{5mm} \par \noindent
         \quad{\it Proof of #1:\hspace{2mm}}}{\endIEEEproof
}

\def\Label#1{\label{#1}\ [\ \text{#1}\ ]\ }
\def\Label{\label}

\begin{document}
\title{Verifiable Quantum Secure Modulo Summation}
%
\author{Masahito~Hayashi\thanks{The first author is with the Graduate School of Mathematics, Nagoya University, Japan. He is also with 
Shenzhen Institute for Quantum Science and Engineering, Southern University of Science and Technology, Shenzhen, China,
Center for Quantum Computing, Peng Cheng Laboratory, Shenzhen, China,
and the Center for Quantum Technologies, National University of Singapore, Singapore.
e-mail:masahito@math.nagoya-u.ac.jp}
and 
Takeshi Koshiba\thanks{The second author is with 
Faculty of Education and Integrated Arts and Sciences, Waseda University, Tokyo, Japan. 
e-mail: tkoshiba@waseda.jp.}
}
\maketitle

\begin{abstract}
We propose 
a new cryptographic task, which we call 
{\em verifiable quantum secure modulo summation}.
Secure modulo summation is a calculation of 
modulo summation 
$Y_1+\ldots+ Y_m$ when 
$m$ players have their individual variables $Y_1,\ldots, Y_m$
with keeping the secrecy of the individual variables.
However, 
the conventional method for secure modulo summation 
uses so many secret communication channels.
We say that a quantum protocol for secure modulo summation is
quantum verifiable secure modulo summation
when it can verify the desired secrecy condition.
If we combine device independent quantum key distribution,
it is possible to verify such secret communication channels.
However, it consumes so many steps.
To resolve this problem,
using quantum systems,
we propose a more direct method to 
realize secure modulo summation with verification.
To realize this protocol, 
we propose modulo zero-sum randomness as another new concept, and show that 
secure modulo summation can be realized by using 
modulo zero-sum randomness.
Then, we construct a verifiable quantum protocol method to generate 
modulo zero-sum randomness.
This protocol can be verified only with minimum requirements.
\end{abstract}

\section{Introduction}
\subsection{Quantum secure modulo summation via
secure modulo zero-sum randomness and its verification}
Secure multiparty computation is an interesting topic in quantum information processing.
Secure modulo summation is a typical example of secure multiparty computation \cite{CK,CS}.
Using secure modulo summation, we can realize visual secret sharing \cite{NS95,KK87}.
In this problem, 
$m$ players have their individual variables $Y_1,\ldots, Y_m$.
The goal of a typical case is 
that all players commonly obtain the modulo summation $Y_1+\ldots+ Y_m$.
As the secrecy condition,
Player $i$'s variable $Y_i$ is not leaked to other players
even when remaining $m-2$ players collude at most, which is referred as the minimum non-collusion condition.
So many existing classical protocols realize this task
by employing so many secret channels. 
If information on all the secret channels is leaked to the third party,
the secrecy of Player $i$'s variable $Y_i$ does not hold.
That is, to guarantee the secrecy,
each player needs to verify all secret channels. 
One natural method for verification of secret channel
is secure communication by using secret key generated by device independent quantum key distribution \cite{Mayers1998,Mayers2004,ABGMPS,PABGMV}. 
However, even when quantum channels are available,
it requires complicated combinations of use of quantum channels dependently on our assumption.
For example, 
if Player $i$ wants to verify the secrecy of the key to be shared by
two other players,
Player $i$ needs to verify the quantum channel between the two players.
For this aim, 
Player $i$ needs to communicate with the two players via secret channels. 
Therefore, 
Player $i$ consumes so many secret keys shared with the two players.
In this way, the verification of conventional methods are not so simple.
Although several quantum protocols for 
secure modulo summation are proposed \cite{SMZ,ZSH,YY,ZRSHS}, their verification has not been discussed.

The aim of this paper is  to provide a construction of a more direct quantum protocol 
for secure modulo summation with verification.
Moreover, we propose modulo zero-sum randomness as another new concept.
In cryptography, we often focus on cryptographic resources
such as secure agreed keys and common randomness:
secure agreed keys play an important role for 
message authentication and common reference strings are essential
for the universal composable security.
Secure modulo zero-sum randomness is a generalization of secure agreed key.
When $m$ players exist,
secure modulo zero-sum randomness is given as random numbers 
$X_i$ in $\FF_2^c$ for $i=1,\ldots, m$ as follows.
The relation $\sum_{i=1}^m X_i=0$ holds and
any $m-1$ variables among $X_1, \ldots, X_m$ are independent of each other.
Player $i$ has the randomness $X_i$ and does not know any other 
random variables except for the above zero-sum condition.
The secure modulo zero-sum randomness is a kind of correlated randomness.
Once $m$ players share secure modulo zero-sum randomness,
using broadcast public channel, i.e., a special channel that cannot be altered nor blocked and can be broadcast to all players,
the $m$ players can realize secure modulo summation.

The big advantage of use of quantum system is selftesting.
Selftesting offers the verification of quantum measurement and states only with the minimum assumption \cite{Mayers1998,Mayers2004,McKague2011,McKague2010,McKague2012,Li,HH2019}.
That is, we do not need to trust any quantum device,
and it is sufficient to assume the independence among several measurement devices.
We propose a quantum protocol to generate 
secure modulo zero-sum randomness as follows.
First, the $m$ players share the GHZ state with respect to the phase basis.
Then, they measure their own system with computation basis.
Since the GHZ state can be regarded as a two-colorable graph state,
it can be verified by selftesting \cite{McKague2011,Li,HH2019}.
However, all players need to verify it without trusting 
other players only with the minimum non-collusion condition.
Due to this requirement, we cannot directly apply 
the existing methods for selftesting of the GHZ state
because they did not care the minimum non-collusion condition.
In this paper, we propose a new selftesting protocol to verify 
the GHZ state under the minimum non-collusion condition.
This protocol is designed so that 
each player can verify a certain secrecy criterion of the generated GHZ state
when $m-2$ remaining players collude at most.
Combining them, we can realize
a quantum protocol for secure modulo summation with verification.

\subsection{Application of secure modulo zero-sum randomness}
Although secure modulo zero-sum randomness
realizes secure modulo summation in the above way
and secure modulo summation can be applied to visual secret sharing \cite{NS95,KK87},
secure modulo zero-sum randomness has many other useful applications as follows.
These applications show usefulness of our verifiable quantum protocol to generate
secure modulo zero-sum randomness.

\begin{itemize}
\item {\bf Application to secret sharing:}
In the standard setting of multi-party secure computation,
many cryptographic protocols require secure communication channels
between any distinct two players \cite{GMW87,BGW88}.
For example, secure multi-party computation for homomorphic functions 
can be realized without honest majority,
but it requires so many secure communication channels \cite{CK}.
Also, any existing secret sharing protocol requires many secure communication 
channels \cite{Shamir,IOS12,PW91,RAX,XMT1,XMT2,Rabin,HK}.
In this paper, using the secure modulo zero-sum randomness, 
we propose protocols to realize these tasks without secure communication 
channels (but broadcast public channel). That is, based on secure modulo zero-sum 
randomness,
we construct a protocol for multi-party secure computation for 
some additively homomorphic functions without honest majority 
nor secure communication channels.
Also, based on the secure modulo zero-sum randomness,
we construct secret sharing protocols without secure communication channels.
We first give a basic protocol for secret sharing
without secure communication channels. Then, by utilizing universal
hash functions, we adapt the basic protocol to a cheater detectable protocol
without secure communication channels.

\item {\bf Application to securely computing additively homomorphic functions:}
A standard method for multi-party secure computation
requires honest majority or secure communication channels.
Another method based on secure message transmission 
\cite{DDWY93,ACdH06,KS09,SZ16}
realizes multi-party secure computation
without honest majority nor secure communication channels.
Instead of secure communication channel, we can employ 
secure message transmission, which is a cryptographic protocol between two parties,
between which there are several channels but some of them are corrupted, 
to send messages 
privately and reliably. 
Secure message transmission protocols can simulate
a secure communication channel between the two parties.
In the standard setting of {\em perfectly} secure message transmission, 
honest majority over the channels is required.
If the broadcast public channel is available in secure message transmission, then
such a barrier can be overcome \cite{FW00,SJST11,KS10} and 
multi-party secure computation can be realized by using secure message
transmission with the broadcast public channel \cite{GO08}. 
However, the respective simulations of the secure communication channels
are quite inefficient.
To resolve this problem,
based on the secure modulo zero-sum randomness,
this paper proposes 
an alternative method for securely computing additively homomorphic functions.
Our protocol uses only 
broadcast public channel as well as 
the secure modulo zero-sum randomness.

\item {\bf Application to multi-party anonymous authentication:}
As another application, we propose {\em multi-party anonymous authentication}, 
which is a new cryptographic task.
Consider the case when a certain project requires the approvals from all the players.
We are required to verify that all the players approve the project by confirming the contents of the project.
Additionally, we might require the anonymity for this approval due to the following reason.
This is because if a person disagreeing to the project can be identified,
a player might hesitate to disagree to it even when he/she does not agree on it in his/her mind.
In this paper, using secure modulo zero-sum randomness,
we construct a protocol to realize multi-party anonymous authentication
without secure communication channel.
\end{itemize}

Indeed, secure modulo zero-sum randomness
can be generated by multi-party secure computation for modulo sum.
In this sense, the generation of secure modulo zero-sum randomness
can be regarded as an equivalent task to multi-party secure computation 
for modulo sum.
In addition, we also discuss several methods to generate 
secure modulo zero-sum randomness.

\subsection{Organization of this paper}
This paper is organized as follows. Section \ref{S2} defines a
new cryptographic resource {\em modulo zero-sum randomness}
and discusses the equivalence to related secure computation protocols.
Section \ref{S7} shows that if we are allowed to use quantum algorithms
it is possible to verify that the resource satisfies the property of
secure modulo zero-sum randomness.
By combining the results in Sections \ref{S2} and \ref{S7},
Section \ref{S8} proposes a quantum verifiable protocol for 
secure modulo summation.
Section \ref{S9} 
compares our method with other methods.
Section \ref{S3} extends the results in Section \ref{S2} to
secure computation with respect to additively homomorphic functions.
Section \ref{S4} provides secret sharing protocols without secure
communication channels. Section \ref{S5} proposes a new cryptographic
task {\em multi-party anonymous authentication}, which employs
modulo zero-sum randomness.
Appendix \ref{S7B} gives 
the generalization to the case with a general finite filed $\FF_q$
when we trust our measurement devices.


\section{Secure Modulo Zero-Sum Randomness}\Label{S2}
First, we give the rigorous definition of secure modulo zero-sum randomness 
for the random numbers $X_i \in \FF_2^c$ with $i=1,\ldots,m$.
The random numbers $X_i \in \FF_2^c$ with $i=1,\ldots,m$
is called {\em secure modulo zero-sum randomness}
when the following conditions hold.

\begin{description}
\item[\rm (1)]
{\it Modulo zero condition}: The relation $\sum_{i=1}^m X_i=0$ holds.

\item[\rm (2)]
{\it Independence condition}:
Any $m-1$ variables among $X_1, \ldots, X_m$ are independent of each other
and subject to the uniform distribution.

\item[\rm (3)]
{\it Secrecy condition}:
Player $i$ has the randomness $X_i$ and does not know any other 
random variables except for the modulo zero condition.
Let $W_i$ be the information of Player $i$ except for $X_i$.
Then, the relation 
$I(X_1, \ldots, X_m; X_i W_i)=
I(X_1, \ldots, X_m; X_i )$ holds.
\end{description}

Using modulo zero-sum randomness, 
we can realize 
the secure calculation of the modulo sum $Y_1+ \cdots + Y_m $ as a function 
with $m$ inputs $Y_i \in \FF_2^c$
without revealing the information for respective inputs.
Here, the $m$ inputs are given by $m$ different players, and 
it is required to calculate the output without informing their inputs to other players.
It is known that secure multi-party computation
for modulo sum is possible without honest majority \cite{CK}.
That is, even when the majority of players do not behave honestly,
the secrecy of each input can be guaranteed.
However, it requires secure communication channels.
When no secure communication channel is available,
to realize the above task only with broadcast public channels, 
it is natural to employ cryptographic resources.

Now let us define the task of secure modulo summation
when player $j$ has the secret input $Y_j \in \FF_2^c$ for $j=1, \ldots, m$.
\begin{description}
\item[\rm (1)]
{\it Reliability condition}: 
Any player $i$ must calculate the modulo sum $Y_1+ \cdots + Y_{m} $
when all players are honest.

\item[\rm (2)]
{\it Secrecy condition}:
Assume that $m-2$ players except for Player $j$ collude at most, which is called the 
{\it minimum non-collusion condition}. 
Also,
their variables $Y_1, \ldots, Y_m$ are assumed to be independent of each other and subject to the uniform distribution.
Then, the variable $X_j$ of player $j$ is independent of 
the information $Z$ obtained by the $m-2$ colluded players. 
That is, the relation 
$I(X_j; Z)=0$ holds.
\end{description}

Indeed, when secure modulo zero-sum randomness $X_1, \ldots, X_m$ is shared,
the following protocol realizes secure modulo summation.

\begin{Protocol}                  
\caption{Secure Modulo Sum Protocol from Secure Modulo Zero-sum Randomness}    
\Label{protocol1}      
\begin{algorithmic}
\STEPONE
Player $i$ sends the information $Z_i:= Y_i+X_i$ to all players via broadcast public channel.

\STEPTWO
Each player calculates 
$\sum_{i=1}^{m} Z_i $, which equals $\sum_{i=1}^m Y_i $.

\end{algorithmic}
\end{Protocol}

\begin{theorem}
Protocol 1 realizes
secure modulo summation.
\end{theorem}

\begin{IEEEproof}
Since Reliability condition holds for the equality
$\sum_{i=1}^{m} Z_i =\sum_{i=1}^m Y_i $,
we show only Secrecy condition.

Due to the symmetry, it is sufficient to show the  
Secrecy condition only for player $1$
when $m-2$ players $3, \ldots, m$ collude.
The $m-2$ players have variables 
$X_3, \ldots, X_m, Y_3, \ldots, Y_m, Z_1,Z_2$.
Since $Y_3, \ldots, Y_m, $ are independent of 
$X_3, \ldots, X_m, X_1+Y_1, X_2+Y_2, Y_1$,
we have
\begin{align}
I(X_3, \ldots, X_m,Y_3, \ldots, Y_m, Z_1,Z_2; Y_1)
=
I(X_3, \ldots, X_m, X_1+Y_1, X_2+Y_2; Y_1).
\end{align}
Since $Y_2$ is subject to the uniform distribution,
$X_3, \ldots, X_m, X_1, X_2+Y_2$
are independent of each other and subject to the uniform distribution.
Hence, 
$H(X_3, \ldots, X_m, X_1+Y_1, X_2+Y_2|Y_1)= m \log q$, which equals
$H(X_3, \ldots, X_m, X_1+Y_1, X_2+Y_2)$.
Thus, 
$I(X_3, \ldots, X_m, X_1+Y_1, X_2+Y_2; Y_1)=0$.
\end{IEEEproof}

Indeed, when secure modulo zero-sum randomness $X_1, \ldots, X_m$ is shared,
the following protocol realizes secure modulo summation.
Further, the discussion in this section can be trivially extended to the case with replacement of $\FF_2$
by $\ZZ_d$ and $\FF_q$.

\begin{Protocol}                  
\caption{Generation of Secure Modulo Zero-sum Randomness from Secure Modulo Summation Protocol}    
\Label{protocol1B}      
\begin{algorithmic}
\STEPONE
Player $i$ generates the variable $Y_i$ subject to the uniform distribution, which is independent of other variables.

\STEPTWO
All players calculate 
the modulo summation $\sum_{i=1}^m Y_i$
by the secure modulo summation protocol.

\STEPTHREE
Player $1$ set the variable 
$X_1:= Y_1-\sum_{i=1}^m Y_i$.
Player $i$ set the variable 
$X_i:= Y_i$ for $i=2, \ldots, m$.

\end{algorithmic}
\end{Protocol}

\begin{theorem}
Protocol \ref{protocol1B} realizes
secure modulo zero-sum randomness.
\end{theorem}

\begin{IEEEproof}
Modulo zero condition follows from 
the relations $\sum_{i=1}^m X_i
= Y_1-\sum_{i=1}^m Y_i+ \sum_{i=2}^m Y_i=0$.

Independence condition holds as follows.
$X_2, \ldots, X_m$ are independent of each other
and subject to the uniform distribution because of their definition.
Next, we focus on $X_1$ and $m-2$ variables among $X_2, \ldots, X_m$. 
As a typical case, we discuss $X_1, \ldots, X_{m-1}$.
Since $Y_m$ is subject to the uniform distribution,
$Y_1-\sum_{i=1}^m Y_i =-\sum_{i=2}^m Y_i$ is also 
subject to the uniform distribution even when $Y_2, \ldots, Y_{m-1}$ are fixed to certain values.
Hence, variables $X_1, \ldots, X_{m-1}$
are independent of each other and subject to the uniform distribution.

Secrecy condition is shown as follows.
To discuss the secrecy of $X_1$, we consider the typical case when Players $2, \ldots, m-1$ collude.
Since $\sum_{i=1}^m Y_i $ is independent of $Y_2, \ldots, Y_{m-1}$,
\begin{align}
I\bigg(X_1;X_2, \ldots, X_{m-2},\sum_{i=1}^m Y_i \bigg)
=&
I\bigg(-\sum_{i=2}^m Y_i ;Y_2, \ldots, Y_{m-1},\sum_{i=1}^m Y_i \bigg)
=
I\bigg(Y_1-\sum_{i=1}^m Y_i  ;Y_2, \ldots, Y_{m-1}\bigg|\sum_{i=1}^m Y_i \bigg)
\nonumber\\
=&
I\bigg(Y_1  ;Y_2, \ldots, Y_{m-2}\bigg|\sum_{i=1}^m Y_i \bigg)
=0.
\end{align}
To discuss the secrecy of $X_2$, we consider the typical case when Players $3, \ldots, m$ collude.
Since $\sum_{i=1}^m Y_i $ is independent of $Y_3, \ldots, Y_{m}$,
\begin{align}
I\bigg(X_2;X_3, \ldots, X_{m},\sum_{i=1}^m Y_i \bigg)
=&
I\bigg(Y_2 ;Y_3, \ldots, Y_{m},\sum_{i=1}^m Y_i \bigg)
=
I\bigg(Y_2  ;Y_1+Y_2+\sum_{i=3}^m Y_i \bigg|Y_3, \ldots, Y_{m}\bigg)
\nonumber\\
=&I\bigg(Y_2  ;Y_1+Y_2\bigg|Y_3, \ldots, Y_{m}\bigg)
=0.
\end{align}
As another case, we consider the case when Players $1, 3, \ldots, m-1$ collude.
Since $ \sum_{i=2}^m Y_i$ and $\sum_{i=1}^m Y_i $ are independent of $Y_3, \ldots, Y_{m-1}$,
\begin{align}
& I\bigg(X_2;X_1,X_3, \ldots, X_{m-1},\sum_{i=1}^m Y_i \bigg)
=
I\bigg(Y_2 ; -\sum_{i=2}^m Y_i,Y_3, \ldots, Y_{m-1},\sum_{i=1}^m Y_i \bigg)
\nonumber\\
=&
I\bigg(Y_2  ;-Y_2-Y_m-\sum_{i=3}^{m-1} Y_i, Y_1+Y_2+Y_m+\sum_{i=3}^{m-1} Y_i \bigg|Y_3, \ldots, Y_{m-1}\bigg)
\nonumber\\
=&
I\bigg(Y_2  ;-Y_2-Y_m, Y_1+Y_2+Y_m \bigg|Y_3, \ldots, Y_{m-1}\bigg)
\nonumber\\
=&
I\bigg(Y_2  ;Y_2+Y_m, Y_1\bigg|Y_3, \ldots, Y_{m-1}\bigg)
=0.
\end{align}

\end{IEEEproof}


\section{Quantum Protocol for Secure Modulo Zero-Sum Randomness}\Label{S7}
Now, we propose a direct verifiable construction by using the GHZ state as follows.
For this aim, we introduce the phase basis state.
The phase basis $\{ |z\rangle_p \}_{z\in\mathbb{F}_2}$ is defined as 
\begin{align*}
|z\rangle_p := \frac{1}{\sqrt{2}} 
\sum_{x\in\mathbb{F}_2} (-1)^{ xz} |x\rangle,
\end{align*}
where 
$ |x\rangle $ expresses the computational basis.

The phase GHZ state $
|GHZ\rangle_p:=
\frac{1}{\sqrt{2}}\sum_{z \in \FF_2} |z,\ldots, z \rangle_{p}$
is calculated as
\begin{align}
|GHZ\rangle_p=
\frac{1}{\sqrt{2^{m-1}}} 
\sum_{x_1,\ldots ,x_m\in\mathbb{F}_2:
x_1+\ldots+x_m=0} 
 |x_1,\ldots ,x_m \rangle .
 \end{align}
When all the players apply the measurement on the computational basis 
to the system whose initial state is $|GHZ\rangle_p$,
Player $i$ obtains the variable $X_i$.
Then, the sum of $m$ outcomes, i.e., $\sum_{i=1}^m X_i$ is zero,
and $m-1$ outcomes are subject to the uniform distribution.
Hence, these outcomes satisfy the conditions of secure modulo zero-sum randomness.
That is, when the initial state is guaranteed to be $|GHZ\rangle_p$,
it is guaranteed that the outcomes are secure modulo zero-sum randomness.

When the $m$ players apply Protocol \ref{protocol1} to the generated 
secure modulo zero-sum randomness,
they can realize secure modulo summation.
To verify its secrecy, 
each player has to verify the generated secure modulo zero-sum randomness.
In secure modulo summation,
when we focus on Player $i$,
we assume that $m-2$ remaining players collude at most.
Hence, 
Player $i$ needs to verify that 
the colluded players has no information with respect to $X_i$
under this assumption.
That is, we assume the following assumption.
\begin{description}
\item[(1)]
Player $i$'s quantum measurement has no correlation with those of other players.
\item[(2)]
Remaining players are divided into two groups $S_1$ and $S_2$.
Both groups are not empty.
There is no correlation between the two groups $S_1$ and $S_2$.
\item[(3)]
Player $i$ does not know 
the separation of remaining players by $S_1$ and $S_2$.
\end{description}

Indeed, when we trust their measurement devices,
we can verify the state $|GHZ\rangle_p$ by using the two projections defined by
\begin{align}
\tilde{P}_1& :=\sum_{x_1,\ldots,x_m:x_1+\ldots+x_m=0} 
| x_1,\ldots, x_m\rangle
\langle  x_1,\ldots, x_m |
=\frac{1}{2}(I+\sZ_{1} \cdots \sZ_{m})\Label{BA1}\\
\tilde{P}_2&:=
\sum_{z}| z, \ldots, z\rangle_{p}
~_{p}\langle z, \ldots, z | 
=\prod_{i;i\neq j} \frac{1}{2}(I+\sX_{j} \sX_{i}),
\Label{BA2}
\end{align}
where $\sZ:=\sum_x (-1)^x|x\rangle \langle x|$ and $\sX:= \sum_x |x+1\rangle \langle x|$.
Here, the subscript of $\sZ$ and $\sX$ expresses the Hilbert space to be acted.

Now, we assume that we prepare $2n+1$ copies. 
Then, we randomly choose $n$ copies and apply the test 
$\tilde{P}_1$.
Also, we randomly choose $n$ copies from the remaining $n+1$ copies and apply the test 
$\tilde{P}_2$.
If these tests are passed, 
the remaining copy can be considered to be close to the true state 
$|GHZ\rangle_p$.
This discussion can be extended to the case with a general finite field $\FF_q$.
Appendix \ref{S7B} gives this generalization with a formal statement of this test.

However, when we cannot trust their measurement devices,
the above method does not work.
We need to employ the method of selftesting \cite{Mayers1998,Mayers2004,McKague2011,McKague2010,McKague2012,Li,HH2019}.
The following is the protocol to generate 
secure modulo zero-sum randomness with the verification by Player $i$.
For this protocol, we prepare the following measurements.
\begin{align}
\sA(k):=(\sX+(-1)^k\sZ)/\sqrt{2}
\end{align}
for $k=0,1$.

If they have the system of $4m n+1$ copies, it can be verified as Protocol \ref{protocol7}.
Here, 
to distinguish the real observable from the ideal observable,
we denote the measured observables by $'$.

\begin{table}[t]
\caption{Measurement for each group}
\Label{meas}
\begin{center}
\begin{tabular}{|c|c|}
\hline
 Group &Measurements    \\
\hline
$k$-th group & $\sZ_j' , \sX_{k}'$ \\
\hline
$m+k$-th group & $\sX_j' , \sX_{k}'$ \\
\hline
$2m+k$-th group & $\sA(0)_j' , \sX_{k}'$ \\
\hline
$3m+k$-th group & $\sA(1)_j' , \sX_{k}'$ \\
\hline
$j$-th group & $\sZ_j' , \sZ_{1}', \ldots, \sZ_{j-1}',\sZ_{j+1}', \ldots, \sZ_m'$ \\
\hline
$m+j$-th group & $\sX_j' ,  \sZ_{1}', \ldots, \sZ_{j-1}',\sZ_{j+1}', \ldots, \sZ_m'$ \\
\hline
$2m+j$-th group & $\sA(0)_j' ,  \sZ_{1}', \ldots, \sZ_{j-1}',\sZ_{j+1}', \ldots, \sZ_m'$ \\
\hline
$3m+j$-th group & $\sA(1)_j' ,  \sZ_{1}', \ldots, \sZ_{j-1}',\sZ_{j+1}', \ldots, \sZ_m'$ \\
\hline
\end{tabular}
\end{center}
Here, $k$ is chosen from $1, \ldots, j-1, j+1 \ldots, m$.
\end{table}

\begin{Protocol}                  
\caption{Verifiable Generation of Secure Modulo Zero-Sum Randomness for Player $j$}    
\Label{protocol7}      
\begin{algorithmic}
\STEPONE
Prepare the system of $4m n+1$ copies.
 
\STEPTWO
Player $i$ randomly divides 
the $4m n+1$ copies into 
$4m+1$ groups such that
the final group is composed of one copy and 
the remaining groups are composed of $n$ copies.

\STEPTHREE
Players apply the measurement to the respective groups except for the final group
as shown in Fig. \ref{meas}.

\STEPFOUR
Players except for Player $j$ send 
their outcomes to Player $j$.
Player $j$ checks the following inequalities for their average values for $k=1, \ldots, j-1, j+1 \ldots, m $.
\begin{align}
&\bA[ \sX_j' \sX_k'  ] \ge 1 -\frac{c_1}{n} , \quad
\bA\Big[ -\sZ_j' \Big(\sum_{l\neq j}\sZ_l' \Big) \Big] \ge 1 -\frac{c_1}{n}  ,\\
&\bA
\Big[ \sA(0)_j' \Big(  \sX_k' - \sum_{l\neq j}\sZ_l' \Big)
 +\sA(1)_j' \Big(  \sX_k' + \sum_{l\neq j}\sZ_l' \Big) \Big] \ge 2\sqrt{2} - \frac{c_1}{\sqrt{n}} .
\end{align}
Here, $\bA$ expresses the average of the observed values with respect to the observables inside of the bracket $[~]$. 
If the above test is passed, 
Player $j$ considers that the remaining copy is close to the phase GHZ state $
|GHZ\rangle_p$.

\STEPFIVE
Each Player $k$ measures the final group with $\sZ$ basis,
and obtain the value $X_k$ for $k=1, \ldots, m $.

\end{algorithmic}
\end{Protocol}

\begin{theorem}\Label{T9}
Assume that players are divided into three distinct non-empty groups
Player $j$, $S_1$, and $S_2$.
When groups $S_1$ and $S_2$ do not collude with each other
and Protocol \ref{protocol7} by Player $j$ is passed,    
with significance level $\alpha$, 
Player $j$ finds that
\begin{align}
\|P_{X_j, E}- P_{X_j} P_E\|_1& \le\frac{c_0}{n^{1/8}}\Label{D3},
\end{align}
where $E$ describes all the information obtained by the group $S_2$
and $c_0$ is a constant dependent on $c_1$ and $\alpha$.
\end{theorem}

\begin{theorem}\Label{T9B}
When all players are honest and Protocol \ref{protocol7} by Player $j$ is passed,    
with significance level $\alpha$, 
Player $j$ finds that the obtained distribution $P_{X_1,\ldots X_m}$ satisfies
\begin{align}
\|P_{X_1,\ldots X_m}- P_{X_1,\ldots X_m|ideal} \|_1& \le\frac{c_0'}{n^{1/8}}\Label{D2},
\end{align}
where
$P_{X_1,\ldots X_m|ideal}$ is the ideal distribution of 
secure modulo zero-sum randomness and
 $c_0'$ is a constant dependent on $\alpha$ and $c_1$.
 \end{theorem}

[Note that the significance level is the maximum passing
probability when malicious Bob sends incorrect states
so that the resultant state $\alpha$ does not satisfy Eqs. \eqref{D3} and \eqref{D2}.] 

\begin{proofof}{Theorem \ref{T9}}
Assume that $S_1$ is composed of 
$j_1, \ldots, j_l$.
We focus on the quantum system of Player $j$
and the quantum system of group $S_1$.
The latter system is spanned by 
the basis 
$$|x\rangle_{S_1}:=
\frac{1}{2^{(l-1)/2}}
\sum_{x_{j_1}, \ldots , x_{j_l}:
x_{j_1}+ \ldots + x_{j_l}=x}
| x_{j_1}\rangle_{j_1} \cdots | x_{j_l}\rangle_{j_l}.$$
It is also spanned by 
$
| z\rangle_{S_1;p} 
:=
| z\rangle_{j_1;p} \cdots | z\rangle_{j_l;p}
=\frac{1}{\sqrt{2}}(|0\rangle_{S_2}+(-1)^z|1\rangle_{S_1})$.
We define 
$\sZ_{S_1}:= |0\rangle_{S_1}~_{S_1}\langle 0|-|1\rangle_{S_1}~_{S_1}\langle 1|$
and 
$\sX_{S_1}:= |0\rangle_{S_1}~_{S_1}\langle 1|+|0\rangle_{S_1}~_{S_1}\langle 1|$.
Similarly, we define 
$\sZ_{S_2}$
and
$\sX_{S_2}$.
While the measurement $\sZ_{S_1}$ can be done by 
the measurement $\sZ_{j_1}, \ldots, \sZ_{j_l}$,
the measurement $\sX_{S_1}$ can be done only by the measurement $\sX_{k}$ for any $k \in S_1$.
The same observation holds for $\sZ_{S_2}$ and $\sZ_{S_2}$.
Therefore, our GHZ $|GHZ\rangle_p$
can be considered as 
$\frac{1}{\sqrt{2}}(
\sum_{z} | z\rangle_{p} | z\rangle_{S_1;p} | z\rangle_{S_2;p} )$.

When one measures $\sZ_{S_2}$, 
obtains the outcome $z$, and applies the unitary $\sX_j^{-z }$,
the resultant state
is the Bell state
$\frac{1}{\sqrt{2}}(\sum_{z} | z\rangle_{j;p} | z\rangle_{S_1;p} )$.
When we measure $\sX_{j}$ and $\sX_{S_1}$ to the system in the state
$\frac{1}{\sqrt{2}}(
\sum_{z} | z\rangle_{j;p} | z\rangle_{S_1;p} | z\rangle_{S_2;p} )$,
the measurement outcome does not depend on the measurement outcome of 
$\sZ_{S_2}$.
Therefore, we can consider that
the measurements on the 
$j_1,j,m+j_1,m+j,2m+j_1,2m+j,3m+j_1,3m+j$-th groups 
can be considered as
the measurement required in Proposition \ref{TH18}.
Now, we denote the real operator on the final group by using $''$.
The real quantum system of Player $j$, the groups $S_1$ and $S_2$ by
${\cal H}_j$, ${\cal H}_{S_1}$, and ${\cal H}_{S_2}$.

Using Proposition \ref{TH18},
we can guarantee, with significance level $\alpha$, that there exist
constant $c_2$
and isometries
$U_j:{\cal H}_j''\to  {\cal H}_j$ 
and
$U_{S_1}:{\cal H}_{S_1}''\to  {\cal H}_{S_1}$ 
such that
\begin{align}
\| U_j \sX''_j U_j^\dagger - \sX_j \| & \le c_2 n^{-1/4}, ~
\| U_j \sZ''_j U_j^\dagger - \sZ_j \|  \le c_2 n^{-1/4}, 
\Label{A6}\\
\| U_{S_1} \sX''_{S_1} U_{S_1}^\dagger - \sX_{S_1} \| & \le c_2 n^{-1/4}, ~
\| U_{S_1} \sZ''_{S_1} U_{S_1}^\dagger - \sZ_{S_1} \|  \le c_2n^{-1/4} .
\Label{A61}
\end{align}
We apply the same discussion to the case with switching $S_1$ and $S_2$.
Then, we can guarantee, with significance level $\alpha$, that there exists
isometry
$U_{S_2}:{\cal H}_{S_2}''\to  {\cal H}_{S_2}$ 
such that
\begin{align}
\| U_{S_2} \sX''_{S_2} U_{S_2}^\dagger - \sX_{S_2} \| & \le c_4 \sqrt{c_1} n^{-1/4}, ~
\| U_{S_2} \sZ''_{S_2} U_{S_2}^\dagger - \sZ_{S_2} \|  \le c_4 \sqrt{c_1} n^{-1/4}.\Label{A62}
\end{align}

We define two projections
\begin{align}
P_1& :=\sum_{x_1,x_2,x_3:x_1+x_2+x_3=0} 
| x_1\rangle_{j}| x_2\rangle_{S_1} | x_3\rangle_{S_2}  
~_{j}\langle x_1 |~_{S_1}\langle x_2 |~_{S_2}\langle x_3 |
\nonumber \\
&=\frac{1}{2}(I+\sZ_{j} \sZ_{S_1} \sZ_{S_2})\Label{B1}\\
P_2&:=
\sum_{z}| z\rangle_{j;p}| z\rangle_{S_1;p} | z\rangle_{S_2;p}  
~_{j;p}\langle z |~_{S_1;p}\langle z |~_{S_2;p}\langle z | \nonumber \\
&=\frac{1}{4}(I+\sX_{j} \sX_{S_1})(I+ \sX_{j}\sZ_{S_2})
=\frac{1}{4}(I+\sX_{j} \sX_{S_1}+ \sX_{j}\sZ_{S_2}+ \sX_{S_1}\sZ_{S_2})
.\Label{B2}
\end{align}
Then, we have 
$|GHZ\rangle_p~_p\langle GHZ|
=P_1 P_2$.
Hence, for $U=U_{j}U_{S_1}U_{S_2}$, 
using \eqref{B1}, we have
\begin{align}
\|U^\dagger P_1U -P_1'' \|
&=\|P_1-U P_1''U^\dagger  \|
\le \frac{1}{2}
(\| \sZ_j-U_j \sZ_j''U_j^\dagger  \|
+\| \sZ_{S_1}-U_{S_1} \sZ_{S_1}''U_{S_1}^\dagger  \|
+\| \sZ_{S_2}-U_{S_2} \sZ_{S_2}''U_{S_2}^\dagger  \|)  \Label{A64}\\
\|U^\dagger P_2 U -P_2'' \|
&=\|P_2-U P_2''U^\dagger  \| 
\le \frac{1}{2}
(\| \sZ_j-U_j \sZ_j''U_j^\dagger  \|
+\| \sZ_{S_1}-U_{S_1} \sZ_{S_1}''U_{S_1}^\dagger  \|
+\| \sZ_{S_2}-U_{S_2} \sZ_{S_2}''U_{S_2}^\dagger  \|) .\Label{A65}
\end{align}

Applying Proposition \ref{TH7} to ${P}''_1$ and ${P}''_2$, 
with significance level $2\alpha$ and a constant $c_2'$,
we have
\begin{align}
\Tr \sigma (I-P''_i) \le \frac{c_2'}{n}\Label{A63}
\end{align}
for $i=1,2$.
Combining \eqref{A6},\eqref{A61}, \eqref{A62}, \eqref{A64}, \eqref{A65}, 
and \eqref{A63}, with significance level $4\alpha$,
we have
\begin{align}
& \Tr \sigma U^\dagger  (I-|GHZ\rangle_p~_p\langle GHZ|)U
\nonumber \\
\le &
\Tr  \sigma U^\dagger ((I-P_1)+(I-P_2))U
\le
\Tr  \sigma ( U^\dagger (I-P_1)U+ U^\dagger (I-P_2)U ) \nonumber \\
\le &
\Tr  \sigma 
( (I-P_1'')+ (I-P_2'') )
+\|U^\dagger P_1U -P_1'' \|
+\|U^\dagger P_2U -P_2'' \| \nonumber \\
\le &
\frac{2c_2'}{n}+3 c_2 n^{-1/4}.
\end{align}
Hence, 
\begin{align}
\| \sigma  -U^\dagger |GHZ\rangle_p~_p\langle GHZ| U\|_1
\le
 \sqrt{\frac{2c_2'}{n}+3 c_2 n^{-1/4}}.
\end{align}

Let 
$\tilde{P}_{X_j,X_{S_1},X_{S_2},E}$ be the joint distribution when 
Players apply the ideal measurements 
$U_j^\dagger  \sZ_jU_j$,
$U_{S_1}^\dagger  \sZ_{S_1}U_{S_1}$, and 
$U_{S_2}^\dagger  \sZ_{S_2}U_{S_2}$.
With significance level $4\alpha$, we have
\begin{align}
& \|P_{X_j,E}-P_{X_j}P_{E}\|_1
\le
\|\tilde{P}_{X_j,E}-\tilde{P}_{X_j}P_{E}\|_1
+\|\tilde{P}_{X_j,E}-P_{X_j,E}\|_1
+\|\tilde{P}_{X_j}-P_{X_j}\|_1 \nonumber \\
\le &
\| \sigma  -U^\dagger |GHZ\rangle_p~_p\langle GHZ| U\|_1
+2\| \sZ_j-U_j \sZ_j''U_j^\dagger  \| \nonumber \\
\le &
 \sqrt{\frac{2c_2'}{n}+3 c_2 n^{-1/4}}
 +2c_2 n^{-1/4}.\Label{C1}
\end{align}
Therefore, with significance level $4\alpha$, we have
\eqref{C1}. 
Hence, replacing $\alpha$ by $\alpha/4$, 
we obtain the desired statement.
\end{proofof}

\begin{proofof}{Theorem \ref{T9B}}
We apply Proposition \ref{TH18} to the case with
$\sX_i$,
$\sX_j$,
$\sZ_i$, and 
$\sZ_j$ for $i \neq j$.
With significance level $\alpha$,
we can guarantee that there exist
a constant $c_2$ and isometries
$U_i:{\cal H}_i''\to  {\cal H}_i$ 
and
$U_j:{\cal H}_j''\to  {\cal H}_i$ 
such that
\begin{align}
\| U_i \sX''_i U_i^\dagger - \sX_i \| \le c_2 n^{-1/4}, ~
\| U_j \sX''_j U_j^\dagger - \sX_j \| \le c_2 n^{-1/4} ,
\| U_i \sZ''_i U_i^\dagger - \sZ_i \|  \le c_2 n^{-1/4}.
\| U_j \sZ''_j U_j^\dagger - \sZ_j \|  \le c_2 n^{-1/4}.
\Label{BA0}
\end{align}
With significant level $(m-1)\alpha$, we have \eqref{BA0} with any $i\neq j$.
Then, using the projections $\tilde{P}_1$ and $\tilde{P}_2$ defined in \eqref{BA1} and \eqref{BA2},
we have 
$|GHZ\rangle_p~_p\langle GHZ|
=\tilde{P}_1 \tilde{P}_2$.
Hence, for $U=U_{j}U_{S_1}U_{S_2}$, 
using \eqref{B1}, we have
\begin{align}
\|U^\dagger \tilde{P}_1U -\tilde{P}_1'' \|
&=\|\tilde{P}_1-U \tilde{P}_1''U^\dagger  \|
\le \frac{1}{2}
\sum_{i=1}^m \| \sZ_i-U_i \sZ_i''U_i^\dagger  \|\Label{BA4}
\\
\|U^\dagger \tilde{P}_2 U -P\tilde{P}_2'' \|
&=\|\tilde{P}_2-U \tilde{P}_2''U^\dagger  \| 
\le \frac{1}{2}
\sum_{i:i\neq j}(\| \sZ_j-U_j \sZ_j''U_j^\dagger  \|
+\| \sZ_{i}-U_{i} \sZ_{i}''U_{i}^\dagger  \|) \nonumber \\
&= 
\frac{m-1}{2}(\| \sZ_j-U_j \sZ_j''U_j^\dagger  \|
+\frac{1}{2}
\sum_{i:i\neq j}
\| \sZ_{i}-U_{i} \sZ_{i}''U_{i}^\dagger  \|)\Label{BA5}
\end{align}

Applying Proposition \ref{TH7} to $\tilde{P}''_1$ and $\tilde{P}''_2$, 
with significance level $2\alpha$ and a constant $c_2'$,
we have
\begin{align}
\Tr \sigma (I-\tilde{P}''_i) \le 
\frac{c_2'}{n}\Label{BA3}
\end{align}
for $i=1,2$.

Combining 
\eqref{BA0},
\eqref{BA4},
\eqref{BA5}, and 
\eqref{BA3},
with significance level $(m+1)\alpha$,
we have
\begin{align}
& \Tr \sigma U^\dagger  (I-|GHZ\rangle_p~_p\langle GHZ|)U
 \nonumber \\
\le &
\Tr  \sigma U^\dagger ((I-\tilde{P}_1)+(I-\tilde{P}_2))U
\le
\Tr  \sigma ( U^\dagger (I-P_1)U+ U^\dagger (I-P_2)U )  \nonumber \\
\le &
\Tr  \sigma 
( (I-\tilde{P}_1'')+ (I-\tilde{P}_2'') )
+\|U^\dagger \tilde{P}_1U -\tilde{P}_1'' \|
+\|U^\dagger \tilde{P}_2U -\tilde{P}_2'' \|  \nonumber \\
\le & 
\frac{2 c_2'}{n}
+\frac{m+2(m-1)}{2} c_2 n^{-1/4}.
\end{align}
Hence, 
\begin{align}
\| \sigma  -U^\dagger |GHZ\rangle_p~_p\langle GHZ| U\|_1
\le
 \sqrt{\frac{2 c_2'}{n}
+\frac{3m-2}{2} c_2 n^{-1/4}
}.
\end{align}

When we apply the measurement based on a POVM  $M=\{M_i\}$ to the system
whose state is $\rho$, 
we denote the output distribution by ${\cal P}_\rho^M$.
For any POVM  $M=\{M_i\}$, 
we have
\begin{align}
& \|{\cal P}_{\sigma}^{M}-{\cal P}_{|GHZ\rangle_p~_p\langle GHZ|}^{M} \|_1
\le
\sum_{i} \Tr M_i | \sigma - |GHZ\rangle_p~_p\langle GHZ||
=
\| \sigma - |GHZ\rangle_p~_p\langle GHZ|\|_1 \nonumber \\
\le &
 \sqrt{\frac{2 c_2'}{n}
+\frac{3m-2}{2} c_2 n^{-1/4}}
.\Label{MM1}
\end{align}

We denote the POVM corresponding to 
the ideal observables $\sZ_1, \ldots, \sZ_m$
(the real observables $\sZ_1'', \ldots, \sZ_m''$)
by $M_{ideal}$ ($M_{real}$).
When we apply the measurement based on the POVM  $M_{ideal}$ ($M_{real}$) to the system
whose state is $\sigma$, 
we denote the output distribution by $P_{X_1,\ldots,X_m}^{M_{ideal}}$
($P_{X_1,\ldots,X_m}^{M_{real}}$).
Since
\begin{align}
P_{X_1,\ldots, X_m}^{M_{real}}
=P_{X_1}^{M_{real}}
P_{X_2|X_1}^{M_{real}}
\cdots 
P_{X_m|X_1,\ldots, X_{m-1}}^{M_{real}}
-P_{X_1,\ldots, X_m}^{M_{ideal}} ,
\end{align}
we have
\begin{align}
& \|{\cal P}_{\sigma}^{M_{real}}-{\cal P}_{\sigma}^{M_{ideal}} \|_1
=
\|P_{X_1,\ldots, X_m}^{M_{real}}-P_{X_1,\ldots, X_m}^{M_{ideal}} \|_1 \nonumber \\
=&
\sum_{i=1}^m\|
P_{X_1}^{M_{real}}
P_{X_2|X_1}^{M_{real}}
\cdots 
P_{X_i|X_1,\ldots, X_{i-1}}^{M_{real}}
\cdots 
P_{X_m|X_1,\ldots, X_{m-1}}^{M_{ideal}}
-
P_{X_1}^{M_{real}}
P_{X_2|X_1}^{M_{real}}
\cdots 
P_{X_i|X_1,\ldots, X_{i-1}}^{M_{ideal}}
\cdots 
P_{X_m|X_1,\ldots, X_{m-1}}^{M_{ideal}}\|_1 \nonumber \\
=&
\sum_{i=1}^m\|
P_{X_1}^{M_{real}}
P_{X_2|X_1}^{M_{real}}
\cdots 
P_{X_i|X_1,\ldots, X_{i-1}}^{M_{real}}
-
P_{X_1}^{M_{real}}
P_{X_2|X_1}^{M_{real}}
\cdots 
P_{X_i|X_1,\ldots, X_{i-1}}^{M_{ideal}}
\|_1  \nonumber \\
=&
\sum_{i=1}^m
\max_{x_1,\ldots,x_{i-1}}
\|P_{X_i|X_1=x_1,\ldots, X_{i-1}=x_{i-1}}^{M_{real}}
-
P_{X_i|X_1=x_{1},\ldots, X_{i-1}=x_{i-1}}^{M_{ideal}}
\|_1  \nonumber \\
=&
\sum_{i=1}^m
\max_{x_1,\ldots,x_{i-1}}
\| U_j \sZ''_j U_j^\dagger - \sZ_j \|
\le m c_2 n^{-1/4}.\Label{MM2}
\end{align}

Since \eqref{BA0} and \eqref{BA3} hold with significance level $(m+1)\alpha$,
combining \eqref{MM1} and \eqref{MM2}, we have
\begin{align}
&\|P_{X_1,\ldots X_m}- P_{X_1,\ldots X_m|ideal} \|_1
 \le
\|{\cal P}_{\sigma}^{M_{real}}
-{\cal P}_{|GHZ\rangle_p~_p\langle GHZ|}^{M_{ideal}} \|_1 \nonumber \\
\le &
\|{\cal P}_{\sigma}^{M_{real}}-{\cal P}_{\sigma}^{M_{ideal}} \|_1
+
\|{\cal P}_{\sigma}^{M_{ideal}}
-{\cal P}_{|GHZ\rangle_p~_p\langle GHZ|}^{M_{ideal}} \|_1
\le
 \sqrt{\frac{2 c_2'}{n}
+\frac{3m-2}{2} c_2 n^{-1/4}}
+m c_2 n^{-1/4}.
\end{align}
Replacing $\alpha$ by $\alpha/(m+1)$, we obtain the desired statement.
\end{proofof}

\section{Quantum Protocol for Secure Modulo Summation}\Label{S8}
Combining the above two methods, the following protocol realizes
secure modulo summation with verification.
\begin{Protocol}                  
\caption{Verifiable Quantum Secure Modulo Summation}    
\Label{protocol8}      
\begin{algorithmic}
\STEPONE 
$m$ players generate $4 m^2 n+1$ copies of the state $|GHZ\rangle_p$.

\STEPTWO 
Each player randomly chooses distinct $4 m n $ copies 
and apply Steps 2, 3, and 4 of Protocol \ref{protocol7}.
If all the tests are passed, they proceed to the next step.

\STEPTHREE
All the players apply the measurement of the computational basis to the remaining one copy.
The outcomes $X_1, \ldots, X_m$ are used as secure modulo zero-sum randomness.

\STEPFOUR 
Player $i$ sends the information $Z_i:= Y_i+X_i$ to all players via broadcast public channel.

\STEPFIVE
Each player calculates 
$\sum_{i=1}^{m} Z_i $, which equals $\sum_{i=1}^m Y_i $.

 \end{algorithmic}
\end{Protocol}

Due to Theorem \ref{T9},
Player $j$ can verify the secrecy of $X_j$
under the minimum non-collusion condition for $j=1, \ldots, m$.
That is, Secrecy condition holds.
Also, Theorem \ref{T9B} guarantees Reliability condition.
Therefore, we can consider that 
Protocol \ref{protocol8} is a verifiable quantum secure modulo summation.

\section{Comparison with other methods}\Label{S9}
\subsection{Other quantum methods for secure modulo summation}
Using quantum systems,
the references \cite{SMZ,ZSH,YY,ZRSHS}
proposed a protocol to securely calculate modulo summation,
which is essentially equivalent to the generation of 
secure modulo zero-sum randomness.
However, they did not propose a method to verify 
the secrecy and the correctness of their computation.
In our method, instead of a direct computation of secure summation,
we propose a method to generate
secure modulo zero-sum randomness
with a protocol (Protocol \ref{protocol7}) to verify 
the secrecy and the correctness. 

For example, the method proposed by the paper \cite{YY} is summarized as follows.
First, $m-1$ players shares the GHZ state $|GHZ\rangle:=
\frac{1}{\sqrt{q}}\sum_{x \in \FF_2} |x,\ldots, x \rangle$.
Second, Player $i$ applies $\sZ^{X_i}$ and sends the system to 
Player $m$, where $\sZ:=
\sum_{x\in\mathbb{F}_2} \omega^{\tr x} 
|x\rangle\langle x|$ and $i$ runs from $1$ to $m-1$.
Third
Player $m$ measures the total system by the basis
$\Big\{\frac{1}{\sqrt{q}} 
\sum_{x\in\mathbb{F}_2} \omega^{-\tr xz} |x,\ldots,x\rangle\Big\}_{x}$.
Finally, 
Player $m$ sends 
the outcome to all other players.

Since the quantum state used in their method
is the GHZ state,
it can be verified in the same way as in Section \ref{S7}.
In this case, 
Player $i$ for $i\neq m$ can
verify the GHZ state in the same way as Protocol \ref{protocol7}.
However, when 
Player $m$ wishes to verify the secrecy of $Y_m$,
the protocol is not so simple.
In this case, in the verification stage,
Player $m$ needs to ask each player to make measurements and send back the outcomes.
Further, these communications need to be secret, which requires additional quantum communication.
Due to this problem,
our method is more efficient than the combination of the method by 
\cite{YY} and the verification given in Protocol \ref{protocol7}.


\subsection{From secure agreed random numbers}\Label{LLG}
We discuss a method to generate secure modulo zero-sum randomness from
secure agreed random numbers.
Secure modulo zero-sum randomness among $m$ players can be generated from several pairs of secure agreed random numbers as follows.
Assume that Player $i$ and Player $i+1$ share
the secret random number $Z_i \in \FF_2^c$.
Also, we assume that Player $m$ and Player 1 share
the secret random number $Z_m \in \FF_2^c$.
Then, Player 1 
puts the random variable $X_1:= Z_1 -Z_m$,
and Player $i$ puts the random variable $X_i:= Z_i -Z_{i-1}$.
The resultant variables $X_1, \ldots, X_m$ satisfy the condition
$\sum_{i=1}^m X_i=0$ and the independence between
any $n-1$ variables of them.

Indeed, secure agreed random numbers can be generated by using 
quantum key distribution.
This method generate secure modulo zero-sum randomness
if all the players are honest.
However, each player does not have a method to verify whether 
other players are honest.
Therefore, even when we apply 
the randomness generated by this method to 
Protocol \ref{protocol1},
the obtained method does not satisfy the condition of 
verifiable secure modulo summation

However, when Player $j$ wishes to verify the generated secure modulo zero-sum randomness,
he/she needs to verify the secrecy of all the secret random numbers.
In this case, these secret random numbers need to be generated by quantum communication with selftesting.
When a part of the secret random number is Player $j$,
the secrecy can be directly verified by Player $j$.
However, when the secret random number is not shared by Player $j$, 
Player $j$ needs to ask both players sharing the secret random number
to make measurement and send the outcome to Player $j$.
The required communication between Player $j$ and each player
should be secret, which required another quantum communication and selftesting.
Therefore, we can say that
our method is more efficient than 
the above method.

\subsection{Asymptotically approximated generation from information theoretical assumption}
Secure modulo zero-sum randomness among $m$ players can be generated from information theoretical assumption
with asymptotically negligible error.
A sequence of random variables $X_{i,n} \in \FF_2^{c_n}$
is called secure modulo zero-sum randomness 
with asymptotically negligible error
when 
\begin{align}
D( 
P_{X_{1,n},\ldots, X_{m,n}},
P_{\tilde{X}_{1,n},\ldots, \tilde{X}_{m,n}})
\to 0
\end{align}
where $D$ is the variational distance and
$\tilde{X}_{1,n},\ldots, \tilde{X}_{m,n} $
is a secure modulo zero-sum randomness among $m$ players.
Here, $\lim_{n \to \infty}\frac{c_n}{n}$ is called the generation rate.

For example, 
secure modulo zero-sum randomness among $m$ players can be generated 
with asymptotically negligible error
when the multiple access channel satisfies a certain condition
and they can use the multiple access channel $n$ times.
The detail construction will be given in \cite{MH}.
Also, with asymptotically negligible error, it can be extracted from 
the $n$-fold independent and identical distribution of 
a certain joint distribution of $m$ random variables 
$Z_1, \ldots, Z_m$ only with broadcast public communication
when the joint distribution satisfies a certain condition \cite{MH}.

However, this method requires noisy classical channel whose noise level is known to all players.
Unfortunately, there is no method to guarantee such a noisy classical channel.
Therefore, 
this method cannot be considered as a protocol to realize verifiable secure modulo summation.

\section{Secure Multi-party Computation of Homomorphic Functions}\Label{S3}
In the discussions in Sections \ref{S3}, \ref{S4}, and \ref{S5},
we assume that $q$ is a power of $2$ because 
the finite field $\FF_q$ can be regarded as an $\ell$-dimensional vector space over $\FF_2$. Hence,
a $c$-dimensional vector space 
over $\FF_q$ can be regarded as $\FF_2^{c \ell}$.
That is,
a secure modulo zero-sum randomness $X_i \in \FF_q^c$ with $i=1, \ldots, m$
is defined as
a secure modulo zero-sum randomness $X_i \in \FF_2^{c\ell}$ with $i=1, \ldots, m$.
However, 
the discussions in Sections \ref{S3}, \ref{S4}, and \ref{S5} can be extended to the case with a general finite field $\FF_q$
because these discussions hold 
by replacing $\FF_2$ with $\FF_q$.

The discussion in the previous section can be extended to a homomorphic function with respect to addition.
Let $f:(\FF_q^c)^m \rightarrow \FF_q^c$ be an additively homomorphic function 
whose value can be determined by a linear combination of inputs.
That is,
\begin{equation} 
f(Y_1,\ldots,Y_m) = \tilde{f}(\alpha_1Y_1 + \cdots + \alpha_mY_m),\Label{Eq1}
\end{equation}
where $\alpha_1,\ldots,\alpha_m$ are all in $\FF_q^c$ and
$\tilde{f}:\FF_q^c \rightarrow \FF_q^c$ is some function.
For the security, we also assume that $f$ is sensitive in the sense that
the image of $f$ distributes uniformly at random when some argument 
is chosen uniformly at random and the other arguments are fixed.

\begin{Protocol}                  
\caption{Secure Computation for an additively homomorphic function $f$}    
\Label{protocol2}      
\begin{algorithmic}
\STEPONE
Player $i$ computes $Z_i:= \tilde{f}(X_i+\alpha_i Y_i)$ and distributes it to
all the other players via public channel.

\STEPTWO
Each player collects all $Z_1,\ldots, Z_m$ and
computes $\sum_{i=1}^m Z_i$.
\end{algorithmic}
\end{Protocol}

The task can be realized in Protocol \ref{protocol2},
which employs secure modulo zero-sum randomness $X_i\in \FF_q^c$ for
$i=1,\ldots,m$ and the broadcast public channel. 
That is, a player sends a message via the
public channel, any other users can receive the same message.
The security is defined in terms of the real/ideal paradigm of the universal
composability \cite{Canetti01,Canetti18}. We will consider the security
in the (${\cal F}_{\sf pub},{\cal F}_{\sf mzsr}$)-hybrid model, where
${\cal F}_{\sf pub}$ is a functionality of the broadcast public channel and
${\cal F}_{\sf mzsr}$ is a functionality of the modulo zero-sum randomness.
So, it is enough to provide simple definitions (without interaction
with the adversary) of the functionalities.

\subsection*{Functionality ${\cal F}_{\sf pub}$ (Simple Form)}
Upon receiving $({\sf Send}, {\it sid}, R, x)$ from Party $i$,
${\cal F}_{\sf pub}$
outputs $({\sf Sent}, {\it sid}, i, R, x)$ to all parties in $R$,
where {\it sid} is a session id and $R$ is a list of receivers of a message $x$.

\subsection*{Functionality ${\cal F}_{\sf mzsr}$ (Simple Form)}
${\cal F}_{\sf mzsr}^{q,c}$ proceeds as follows, when parameterized by
the alphabet size $q$ and the length $c$.

Upon receiving $({\sf Request}, {\it sid})$ from Party $i$,
${\cal F}_{\sf mzsr}^{q,c}$ generates modulo zero-sum randomness $X_1,\ldots, X_m$, 
each in $\mathbb{F}_2^c$,
satisfying the modulo zero-sum condition and the independence condition
and outputs $({\sf Response}, {\it sid}, i, j, X_j)$ to Party $j$ for each
$j=1,\ldots,m$.

\medskip

Now, we are ready to give definitions of the correctness and the privacy.

\begin{definition}\rm
Let $f(x_1,\ldots,x_m)$ be an $m$-party functionality and $\pi$ be an
$m$-party protocol. We say that the protocol is {\em correct} if
honest parties do not get incorrect values in the presence of the adversary.
\end{definition}

\noindent
{\it Remark.} If $\pi$ is correct, then the following holds.
\begin{enumerate}
\item The protocol aborts whenever it detects a cheating behavior of the
adversary, or
\item honest parties must get the correct values if $\pi$ does not abort.
\end{enumerate}

\begin{definition}\rm
Let $f(x_1,\ldots,x_m)$ be an $m$-party functionality and $\pi$ be an
$m$-party protocol. We say that the protocol $\tau$-{\em securely computes} 
$f$ with {\em perfect privacy} if there exists a simulator $\cal S$ for which the following holds.
For any subset of corrupted parties $T\subseteq \{1,\ldots,m\}$ at most size $\tau$ by the adversary $\cal A$ and every $m$-tuple of inputs $\bm{x}=(x_1,\ldots,x_m)$, two probability distributions 
${\sf Ideal}_{f,\cal S}(\bm{x})$ and ${\sf Real}_{\pi,{\cal A}}$ are
identical.
\begin{enumerate}
\item ${\sf Ideal}_{f,\cal S}(\bm{x})$ is defined as 
\[ ({\cal S}(T,\bm{x}[T],\bm{y}[T]), \bm{y}[\bar{T}]), \]
where $\bm{y}=f(\bm{x})$, $\bar{T}=\{1,\ldots,m\}\setminus T$, and $\bm{v}[T]$ denotes the sub-vector $(v_j)_{j\in T}$
for a vector $\bm{v}=(v_1,\ldots,v_m)$.
This is the joint distribution of the simulated view of the corrupted parties
together with outputs of the honest parties in an ideal implementation of $f$.
\item
${\sf Real}_{\pi,{\cal A}}$ is defined as 
\[({\sf View}_{\pi,T}(\bm{x}), {\sf Output}_{\pi,\bar{T}}(\bm{x})), \]
where ${\sf View}_{\pi,T}(\bm{x})$ is the joint view of the parties in $T$
by executing $\pi$ on input $\bm{x}$ and ${\sf Output}_{\pi,\bar{T}}(\bm{x})$
is the output that $\pi$ delivers to the honest parties in $\bar{T}$. 
\end{enumerate}
\end{definition}

\noindent
{\it Remark.} If $f$ is a single-valued function and the functional value
is required to be shared among all the parties, then we consider that
$\bm{y}=(y,y,\ldots,y)$, where $y=f(\bm{x})$.

\begin{theorem}
Let $f$ be an additively homomorphic function of the form Eq.(\ref{Eq1}).
Then, Protocol \ref{protocol2} is correct in the semi-honest model.
\end{theorem}
\begin{IEEEproof}
Since the adversary does not alter $Z_i$ due to the condition
of broadcast public channel, each player can collect
the correct values $Z_1,\ldots, Z_m$. Then each player
computes
\[ 
\sum_{i=1}^m Z_i =
\sum_{i=1}^m \tilde{f}(X_i+\alpha_i Y_i) = 
\tilde{f}\left(\sum_{i=1}^m X_i + \sum_{i=1}^m 
\alpha_i Y_i\right) = f(Y_1,\ldots,Y_m).
\]
This concludes the proof.
\end{IEEEproof}

\noindent
{\it Remark.} If we allow a malicious adversary $\cal A$, which can send a fake value for $Z_i$ for Party $i$ corrupted by $\cal A$. In this case, 
the correctness
of Protocol \ref{protocol2} does not hold. A naive application of
universal hash functions, which will be discussed in Section \ref{S4},
does not work. In this paper, our concern is to demonstrate that
the modulo zero-sum randomness contributes to simple cryptographic construction
and thus we do not consider the correctness in the malicious model any more.  

\begin{theorem}\Label{Tsc}
Let $f$ be an additively homomorphic function of the form Eq.(\ref{Eq1}).
Then, $f$ can be $(m-1)$-securely computed with perfect privacy by
Protocol \ref{protocol2} 
in the (${\cal F}_{\sf pub},{\cal F}_{\sf mzsr}$)-hybrid model.
\end{theorem}
\begin{IEEEproof}
For the proof, we follow the convention in \cite{Goldreich-book}.
First,  we assume that the adversary $\cal A$ collapses
Players $1,\ldots, m-1$. Since Protocol \ref{protocol2} is essentially
non-interactive, what the adversary $\cal A$ can do is just sending a fake value
$Z_i'$ instead of $Z_i$ for Player $i$.
Then, ${\sf Real}_{\pi,\cal A}$ (with help of ${\cal F}_{\sf pub}$
and ${\cal F}_{\sf mzsr}$) is described as
\[ \{X_1,\ldots, X_{m-1}, Y_1,\ldots, Y_{m-1}, Z_1', \ldots, Z_{m-1}', Z_m,
f(Y_1,\ldots,Y_m) \},
\]
since $\cal A$ can compute $Z_1,\ldots, Z_{m-1}$ and also $f(Y_1,\ldots,Y_m)$
from $Z_1,\ldots,Z_m$.
Now, we construct a simulator $\cal S$ which takes $X_1,\ldots, X_{m-1}$, 
$Y_1, \ldots, Y_{m-1}$ as input with help of ${\cal F}_{\sf mzsr}$.
$\cal S$ can compute $Z_1',\ldots, Z_{m-1}'$ as $\cal A$ does.
Also $\cal S$ can send $Y_1,\ldots, Y_{m-1}$ to the functionality $f$ to
get $f(Y_1,\ldots,Y_m)$. Then $\cal S$ can
compute $Z_m$ as
\[ Z_m = f(Y_1,\ldots, Y_m) - \sum_{i=1}^{m-1}\tilde{f}(X_i + \alpha_i Y_i). \]
Thus, we can say that ${\sf Ideal}_{f,\cal S}$ is identical to ${\sf Real}_{\pi,\cal A}$.

Next, we consider the case that $\cal A$ collapses Players $1,\ldots, k$,
where $k<m-1$. In this case, we can similarly construct a simulator $\cal S$.
The difference is that $\cal S$ can compute $Z = Z_{k+1}+ \cdots + Z_m$ 
instead of $Z_m$. Since $\tilde{f}$ is sensitive, we can take
random values from the image of $\tilde{f}$ for $Z_{k+1},\ldots, Z_{m-1}$.
$\cal S$ can set $Z_m = Z - (Z_{k+1} + \cdots + Z_{m-1})$. 
This is also a perfect simulation of ${\sf Real}_{\pi,\cal A}$.
\end{IEEEproof}

\noindent
{\it Remark.} 
In the statement of Theorem \ref{Tsc}, we do not clearly mention that
the corruption is static or adapive. Since Protocol \ref{protocol2} 
is essentially non-interactive, we do not distinguish static adversaries
from adaptive ones.

\section{Secret Sharing without Secure Communication Channel}\Label{S4}
\subsection{Basic Protocol}
While there are many secret sharing protocols,
they require secure communication channel in the dealing phase \cite{Shamir}.
Now, we propose a secret sharing protocol without use of secure communication channel.
Assume that 
there are $m$ players and Player $1$ has a secret message $ Y \in \FF_q^c$.
Our task is the following without use of secure communication channel.
Player $m$ can decode the secret message $ Y $ only when
all the $m-1$ players except for Player $1$ collaborate for the decoding.
A conventional secret sharing protocol does not achieve this requirement 
because it employs secure communication channels in the dealing step.

When the $m$ players have secure modulo zero-sum randomness 
$X_i \in \FF_q^c$ for $i=1,\ldots,m$,
this task can be realized as Protocol \ref{protocol3}.

\begin{Protocol}                  
\caption{Secret Sharing without secure communication channel}    
\Label{protocol3}      
\begin{algorithmic}
\STEPONE[Dealing]  
Player $1$ sends the information $Z:= X_1+ Y$ to Player $m$ via broadcast public channel.

\STEPTWO
Players $2, \ldots, m-1$ send their randomness $X_2, \ldots, X_{m-1}$ to Player $m$ via broadcast public channel.

\STEPTHREE[Reconstruction]  
Player $m$ reconstructs the original information 
$Z+\sum_{i=2}^m X_i $, which equals $Y$.

\end{algorithmic}
\end{Protocol}

\subsection{Cheater Detectable Protocol}
However, this protocol cannot detect whether 
Players $2, \ldots, m-1$ send incorrect information.
To resolve this problem, we propose the following protocol (Protocol \ref{protocol4}), which employs
secure modulo zero-sum randomness $X_i \in \FF_q^c$ for $i=1,\ldots,m$.
In this protocol, the information $Y$ transmitted from Player $1$ is a non-zero element of $\FF_q$.
Hence, $Y$ is subject to the uniform distribution on $\FF_q\setminus \{0\}$.
When the size of information to be transmitted is large, we use algebraic extension.
We identify the vector space $\FF_q^c$ with the finite filed $\FF_{q'}$ with $q'=q^c$ by considering algebraic extension. 

\begin{Protocol}                  
\caption{Cheater Detectable Secret Sharing without secure communication channel}    
\Label{protocol4}      
\begin{algorithmic}
\STEPONE
Players $2, \ldots, m-1$ send their randomness $X_2, \ldots, X_{m-1}$ to Player $m$ via broadcast public channel.

\STEPTWO[Dealing]
Player $1$ sends the information $Z:= X_1 Y$ to Player $m$ 
via broadcast public channel.

\STEPTHREE[Reconstruction]  
If $Z\neq 0$,
Player $m$ defines
$Y':= -Z(\sum_{i=2}^m X_i)^{-1} $.
If $Y'$  belongs to $\FF_q \subset \FF_{q'}$,
Player $m$ considers that there is no cheating and $Y'$ equals the original information $Y$.
If $Y'$ does not belong to $\FF_q \subset \FF_{q'}$,
Player $m$ considers that there is cheating and discard $Y'$.

\end{algorithmic}
\end{Protocol}

Now, we analyze the performance of Protocol \ref{protocol4}.
First of all, we consider the security of Protocol \ref{protocol4} and
the success probability of the reconstruction
of Protocol \ref{protocol4} 
when all the players are honest.
\begin{proposition}\Label{prop:p4}
Suppose that all the players are honest in Protocol \ref{protocol4}. Then, 
Protocol \ref{protocol4} has the perfect secrecy and
the success probability of the reconstruction 
is $1-q^{-c}$.
\end{proposition}
\begin{IEEEproof}
If $X_1=0$ then $Z=0$. In this case, $Y'=0$. This is different
from $Y$, which is non-zero. If $X_1\ne 0$ then $Y'=Y$ and the
reconstruction succeeds. Since the probability that $X_1=0$ is 
$q^{-c}$, the success probability is $1-q^{-c}$.
For the security, we assume that Players $2,\ldots,m$ collude
to get $X_2+\cdots + X_{m-1}$. If $X_1=0$ then $Z=0$. $Z$ does not include
any information on $Y$. If $X_1\ne 0$ then $Z$ looks random. What they
can do for guessing $X_1$ is just a random choice. This implies
that $X_m$ also looks random. Since $Y=Y'=-Z(\sum_{i=2}^m X_i)^{-1}$
and both its numerator and denominator are non-zero, what they can
do for guessing $Y$ is also a random choice. Thus, 
Protocol \ref{protocol4} has the perfect secrecy.
\end{IEEEproof}

If Players $2, \ldots, m-1$ use the information in the dealing phase,
these players can make a cheat.
Hence, it is essential to put the transmission of the random variables 
$X_2, \ldots, X_{m-1}$ before the dealing phase.
%
As an attack, we assume that at least one of Players $2, \ldots, m-1$ 
makes Player $m$ to decode a different information from $Y$ that belongs to $\FF_q$.
We call this attack the {\em modification attack}.
For simplicity, we consider the case when all of Players $2, \ldots, m-1$ collude for the modification attack.

\begin{theorem}\Label{Thm:md}
When all of Players $2, \ldots, m-1$ in Protocol \ref{protocol4} collude for the modification attack,
they succeed the attack with probability $\frac{q-1}{q'-1}$.
\end{theorem}

\begin{IEEEproof}
When $X_1 \neq 0$, 
to succeed this attack,
the sum $V'$ of variables sent from Players $2, \ldots, m-1$ to $m$ needs to satisfy the 
condition $-X_1^{-1} (V'+X_m) \in \FF_q\setminus \{1\}$.
When we denote the sum $\sum_{i=2}^{m-1}X_i$ by $V$,
the above condition is equivalent to the following condition.
There exists an element $A(\neq 1) \in \FF_q$ such that
$ V'-V-X_1= -A X_1$, i.e., $V'= V+(1-A)X_1$.
Since $X_1$ is 
subject to the uniform distribution on $\FF_{q'} \setminus \{0\}$,
the variable $(1-A)X_1$ is 
subject to the uniform distribution on $\FF_{q'} \setminus \{0\}$.
Since the number of $A(\neq 1) \in \FF_q$ is $q-1$,
the probability to satisfy the condition required to $V'$
is $\frac{q-1}{q'-1}$.
\end{IEEEproof}

\noindent
{\it Remark.} Regardless of the modification attack (in the setting of Theorem \ref{Thm:md}), Protocol \ref{protocol4} maintains the perfect secrecy
as discussed in the proof of Proposition \ref{prop:p4}.

\medskip

Indeed, there exist so many secret sharing protocols with dishonest players.
Some of them can identify the cheating players \cite{Rabin,IOS12,XMT1,RAX,XMT2,PW91,HK}.
However, all the existing protocols require secure communication channels in the dealing phase.
The advantage of this protocol is 
unnecessity of secure communication channels
due to use of secure modulo zero-sum randomness.

\section{Multi-party Anonymous Authentication}\Label{S5}
\subsection{Basic Protocol}
Suppose that a certain project written as the variable $Y \in \FF_q^d $
requires the approvals from all of $m$ players.
Our requirement is the following. We verify that all $m$ players approve the project by confirming the contents $Y$.
Additionally, we require anonymity for this approval.

We consider the following naive protocol by using secure modulo zero-sum randomness 
$X_i \in \FF_q^c$ for $i=1,\ldots,m$.
If Player $i$ agrees on the project, 
he/she sends his/her random variable $X_i$ to the other players via broadcast public channel.
Otherwise, he/she sends another variable to the other players via broadcast public channel.
Then, each player calculates the sum of the received variables and his/her own variable.
If the sum is zero, the project can be considered to be approved.

However, this protocol has the following problem.
There is a possibility that Player $i$ incorrectly receives a different information $Y'$ from $Y$ as the project.
This case is called a {\em mismatched recognition}.
In fact, when the secrecy of the information $Y$ is required,
it might be distributed via secure communication channel priorly.
This assumption is natural because it is usual to require the secrecy of the contents of the project.
Hence, we need to be careful about a mismatched recognition.
That is, we need to verify that each player makes the decision based on the correct information $Y$.

To prevent a mismatched recognition,
as illustrated in Fig. \ref{F1},
attaching the message authentication protocol \cite{K95,M96} 
to information $Y$,
we propose the following protocol as Protocol \ref{protocol5}.
As a preparation of Protocol \ref{protocol5}, 
from secure modulo zero-sum randomness $X_i \in \FF_q^c$ for $i=1,\ldots,m$,
we generate an $e \times d$ Toeplitz matrix $T_i$ and a variable $A_i \in \FF_q^e$,
where we choose the integers $e$ and $d$ to satisfy $2e+d-1=c$.
(Note that Toeplitz matrices can be used universal hash functions.
You may consult with a textbook \cite{toeplitz}.)
Indeed, since an $e \times d$ Toeplitz matrix $T_i$ needs 
$e+d-1$ elements of $\FF_q$,
the pair of $T_i$ and $A_i$ requires 
$2e+d-1=c$ elements of $\FF_q$.
In the following, we also assume that 
the variable $Y \in \FF_q^d $ describing the project has been distributed to all the players priorly while there is a possibility of a mismatched recognition.

\begin{figure*}
\begin{center}
\scalebox{1}{\includegraphics[scale=0.4]{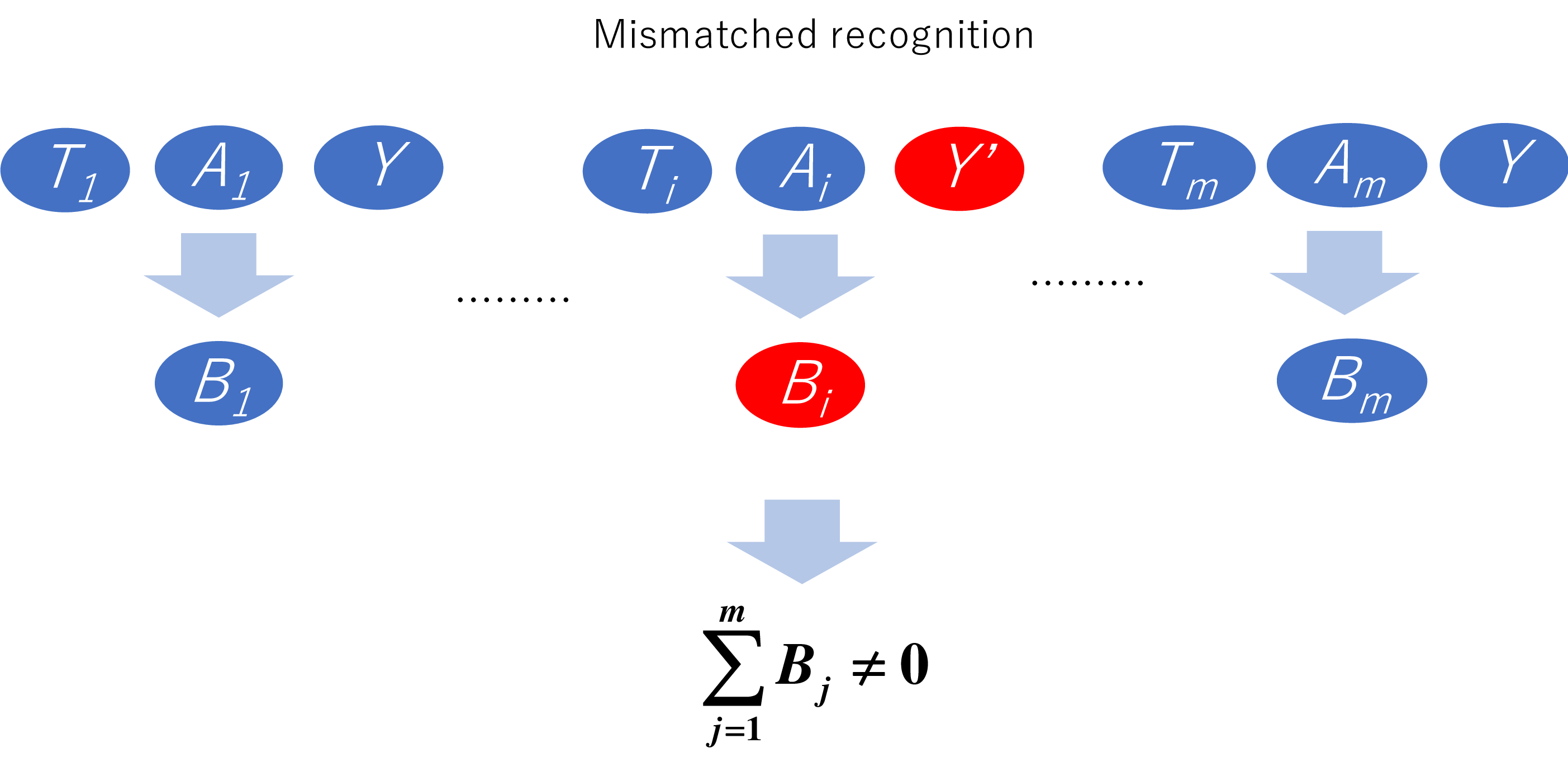}}
\end{center}
\caption{Mismatched recognition.}
\Label{F1}
\end{figure*}%

\begin{Protocol}                  
\caption{Multi-party Anonymous Authentication}    
\Label{protocol5}      
\begin{algorithmic}
\STEPONE[Voting]  
Player $i$ sends $B_i \in \FF_q^{e}$ 
to the remaining players via broadcast public channel.
If Player $i$ agrees on the project described by $Y$,
he/she chooses $B_i$ as $T_i Y + A_i $.
Otherwise, he/she chooses $B_i$ subject to the uniform distribution
on $\FF_q^{e}$.

\STEPTWO[Verification]  
Each player calculates $\sum_{i=1}^n B_i$.
If the sum is zero, the project can be considered to be approved.

\end{algorithmic}
\end{Protocol}

\subsection{Analysis with honest players}
When all the players send $T_i Y + A_i $ based on the same variable $Y$,
we have 
$\sum_{i=1}^n B_i 
=\sum_{i=1}^n T_i Y + A_i 
=(\sum_{i=1}^n T_i) Y + (\sum_{i=1}^n A_i) =0 Y+0=0$
and all the players find that all of them approve the project written by $Y$.
Hence, for security analysis, 
we need the analysis on the case when at least one player disagrees on the project
and/or at least one player recognizes a different information from $Y$.
For this aim, we have the following two theorems.

\begin{theorem}\Label{T1}
When at least one Player $i'$ disagrees on the project, 
the probability of $\sum_{i=1}^n B_i =0$ is $q^{- e}$.
\end{theorem}

\begin{IEEEproof}
Since $B_{i'}$ is subject to the uniform distribution on $\FF_q^{e}$,
the probability of $\sum_{i=1}^n B_i =0$ is $q^{- e}$.
\end{IEEEproof}

\begin{theorem}\Label{T2}
When all the players agree on the project and
at least one Player $i$ recognizes the information $Y_i$ 
that is different from the information $Y_1$ recognized by Player $1$,
the probability of $\sum_{i=1}^n B_i =0$ is $q^{- e}$.
\end{theorem}

This theorem ensures that if the project is approved by this protocol,
all the players confirm no mismatched recognition.

\begin{IEEEproof}
Assume that players $i_1, \ldots, i_k$ recognize the information $Y_{i_1}, \ldots, Y_{i_k}$ 
that is different from the information $Y_1$ recognized by Player $1$.
Also assume that other players recognize the same information $Y_1$ recognized by Player $1$.
We define the variable $V_{i_j}:= Y_{i_j}-Y_1$ for $j=1, \ldots, k$.
Then, we have
\begin{align}
\sum_{i=1}^n B_i = \sum_{j=1}^k T_{i_j} V_{i,j}.
\end{align}
Since $V_{i,j} \neq 0$,
the variable $T_{i_j} V_{i,j}$ is independently subject to the uniform distribution on $\FF_q^{e} $.
Hence, $\sum_{j=1}^k T_{i_j} V_{i,j}$ is also 
subject to the uniform distribution on $\FF_q^{e} $.
Therefore, we obtain the desired statement.
\end{IEEEproof}

\subsection{Analysis with malicious player}
Now, we consider the case with a malicious player.
When malicious Player $j$ makes {\em rushing},
Player $i$ can realize the situation $\sum_{i=1}^n B_i =0 $ 
by sending $-\sum_{i\neq j} B_i  $
unless all the players do not approve the same variable $Y$.
Hence, when we employ Protocol \ref{protocol5},
we need to trust all the players.
To avoid the rushing attack, we propose another protocol
(Protocol \ref{protocol6}), which trusts Player $1$.

 \begin{Protocol}                  
\caption{Secure Multi-party Anonymous Authentication}    
\Label{protocol6}      
\begin{algorithmic}
\STEPONE[Voting]  
Player $i$ sends $B_i \in \FF_q^{e}$ to Player $1$ via broadcast public channel.
If Player $i$ agrees on the project described by $Y$,
he/she chooses $B_i$ as $T_i Y + A_i $.
Otherwise, he/she chooses $B_i$ subject to the uniform distribution
on $\FF_q^{e}$.

\STEPTWO[Verification]  
Player $1$ calculates $\sum_{i=1}^n B_i$, where $B_1:=T_1 Y + A_1$.
If it is zero, the project can be considered to be approved.

\STEPTHREE[Notification]  
Player $1$ sends the above result to other players.
 
\end{algorithmic}
\end{Protocol}

Now, we consider the following type of malicious player.
Assume that malicious Players $l, \ldots, m$ want to make the following situation by colluding together.
Here, for the notational convenience, we assume that 
Players $l, \ldots, m$ are malicious.
Players $1,\ldots, l-1$ consider that the project is described by another information $Y_{1}, \ldots, Y_{l-1}$,
and they approve this project based on this incorrect information.
Then, Player $1$ announces that all the players approve the project based on the same information while they are not the same.
For this kind of attack, we have the following theorem.
\begin{theorem}
In Protocol \ref{protocol6},
malicious Players $l, \ldots, m$ succeed the above attack with probability
$q^{-e}$.
\end{theorem}

\begin{IEEEproof}
To realize this situation, $\sum_{i=l}^m B_i$ needs to be
$-\sum_{i=1}^{l-1} T_i Y_i +A_i$, which is calculated as
\begin{align*}
-\sum_{i=1}^{l-1} T_i Y_i +A_i
=&-\sum_{i=2}^{l-1} T_i (Y_i-Y_1) 
-\sum_{i=1}^{l-1} T_i Y_1+A_i
\\
=&-\sum_{i=2}^{l-1} T_i (Y_i-Y_1) 
+\sum_{j=l}^m T_j Y_1+A_j.
\end{align*}
We define the set $ \{ i_1, \ldots, i_k\}:=
\{ i \in [2,l-1] | Y_i\neq Y_1\}$.
Then, $T_{i_1} (Y_{i_1}-Y_1), \ldots, T_{i_k} (Y_{i_k}-Y_1)$ are independently subject to the uniform distribution on $\FF_q^{e}$.
Since 
$A_i$ is subject to the uniform distribution on $\FF_q^{e}$ for $i=2, \ldots, l-1$,
$B_i $ is independent of $ T_i Y_i $.
Hence, Players $l, \ldots, m$ cannot obtain any information $T_i Y_i $ from $B_i$ for $2, \ldots, l-1$.
Thus, letting 
\begin{align*}
V:=
(B_2, \ldots, B_{m-1}, T_l,\ldots,T_m ,A_l, \ldots, A_m, Y_1, \ldots, Y_{l-1}),
\end{align*}
we obtain
\begin{align}
&I( -\sum_{i=2}^{l-1} T_i (Y_i-Y_1) 
+\sum_{j=l}^m T_j Y_1+A_j ;V)
\nonumber \\
=&I( -\sum_{i=2}^{l-1} T_i (Y_i-Y_1);V) 
=0.
\end{align}
Since $-\sum_{i=2}^{m-1} T_i (Y_i-Y_1)$ is subject to the uniform distribution,
Players $l, \ldots, m$ can make the situation
$\sum_{i=1}^n B_i =0$ with probability $q^{-e}$.
\end{IEEEproof}

\section{Conclusion}
We have proposed a new concept of secure modulo sum randomness
and a quantum protocol to generate it.
We also have constructed its verification protocol that works even with untrusted measurement devices.
Then, combining them, we have proposed 
a verifiable quantum protocol for secure modulo summation for $m$ players
even with untrusted measurement devices.
This protocol guarantees secrecy for each player even when 
$m-2$ players collude at most.
However, since we employ selftesting, 
our method works only with $\FF_2$.
In order to extend our method to the case with a general finite field $\FF_q$,
we need to develop selftesting in $q$-dimensional system with operators 
$\sX$ and $\sZ$.
This is an interesting future study.

\section*{Acknowledgments}
MH is supported in part by a JSPS Grant-in-Aids for Scientific Research (A) No.17H01280 and for Scientific Research (B) No.16KT0017, 
and Kayamori Foundation of Information Science Advancement.
TK is supported in part by a JSPS Grant-in-Aids for Scientific Research (A)
No.16H01705,  for Scientific Research (B) No.17H01695, and for Challenging
Exploratory Research No.19K22849 and MEXT Quantum Leap Flagship Program
(MEXT Q-LEAP) Grant No. JPMXS0118067285.

\appendices

\section{Classical random sampling}
We consider $n+1$ binary random variables $X_1, \ldots, X_{n+1}$ taking values in $\{0,1\}$.
We randomly choose $n$ variables among $X_1, \ldots, X_{n+1}$
and observe them.
We denote the remaining variable by $Y$.
Let $Z$ be the number of $1$ among observed variables.

Lemma 3 of \cite[Appendix C]{HH2019} is rewritten as follows.
\begin{proposition}\Label{TH6}
With significance level $\alpha$, we have
For any constants $c_1$, $p*$ and $\alpha$, there exists a constant $c_2$ such that
with significance level $\alpha$, we have
\begin{align}
p_* - \frac{c_2}{\sqrt{n}}
\le
Pr\Big( Y =1 \Big|   p_* - \frac{c_1}{\sqrt{n}}  \le \frac{Z}{n}\le p_* + \frac{c_1}{\sqrt{n}}\Big) 
\le 
p_* + \frac{c_2}{\sqrt{n}}
\end{align}
\end{proposition}

When $p_*$ is zero, we prepare a different type of evaluation as follows.

\begin{proposition}\Label{TH7}
With significance level $\alpha\ge \frac{k+1}{n+1}$, we have
\begin{align}
Pr( Y =1| Z\le k) \le 
\frac{k}{ \alpha(n+1)}+\frac{1 -\alpha }{ \alpha(n -k)}.
\end{align}
That is, for any constants $c_1$ and $\alpha$, there exists a constant $c_2$ such that
with significance level $\alpha$, we have
\begin{align}
Pr \Big( Y =1 \Big| \frac{Z}{n}\le \frac{c_1}{n} \Big)\le 
\frac{c_2}{n}
\end{align}
\end{proposition}

\begin{IEEEproof}
We denote the number of $1$ among $X_1, \ldots, X_{n+1}$
by the variable $X$.
We assume that $P(X=x)=P_x$.
Then, we have 
\begin{align}
Pr(Z=z,Y=y) =
P_{z+y}
\frac{{n \choose z}}{{n+1 \choose z+y}}.
\end{align}
That is,
\begin{align}
Pr(Z=z,Y=0) &=P_{z} \frac{n -z+1}{n+1 } \\
Pr(Z=z,Y=1) &=P_{z+1} \frac{z+1}{n+1}
\end{align}
Thus, we have
\begin{align}
Pr(Z=z) =
P_{z}\frac{n -z+1}{n+1 }
+P_{z+1} \frac{z+1}{n+1}.
\end{align}
Hence, 
\begin{align}
& Pr(Z \le k, Y=1)
= \sum_{z=0}^k 
P_{z+1} \frac{z+1}{n+1},
\end{align}
and
\begin{align}
\frac{Pr(Z\le k,Y=1) }{Pr(Z\le k) }=
\frac
{\sum_{z=0}^k P_{z+1} \frac{z+1}{n+1}}
{(\sum_{z=0}^k P_{z} )
+P_{k+1} \frac{k+1}{n+1}}
\end{align}

Since $\frac{1}{n+1}\le \ldots \le \frac{k}{n+1} \le \frac{k+1}{n+1}$
and $1 \ge \frac{k+1}{n+1}$,
we have 
\begin{align}
&\max_{(P_z)_{z=0}^{n+1}}
\Big\{
\frac
{\sum_{z=0}^k P_{z+1} \frac{z+1}{n+1}}
{(\sum_{z=0}^k P_{z} )
+P_{k+1} \frac{k+1}{n+1}}
\Big|
(\sum_{z=0}^k P_{z} )
+P_{k+1} \frac{k+1}{n+1} \ge \alpha
\Big\}
\nonumber \\
&=\max_{p}
\Big\{
\frac
{(1-p) \frac{k}{n+1}+p \frac{k+1}{n+1}}
{(1-p)+p \frac{k+1}{n+1}}
\Big|
(1-p)+p \frac{k+1}{n+1}
\ge \alpha
\Big\}\Label{L2}
\end{align}

The condition $(1-p)  
+p \frac{k+1}{n+1}
\ge \alpha$ is equivalent to 
the condition 
$  1 -\alpha
\ge 
p( 1- \frac{k+1}{n+1})
=
p \frac{n -k}{n+1 }$, which is rewritten as
$  (1 -\alpha)\frac{n+1 }{n -k}
\ge p$.
Under this condition, we have
\begin{align}
&\frac
{(1-p) \frac{k}{n+1}+p \frac{k+1}{n+1}}
{(1-p)+p \frac{k+1}{n+1}}
=\frac
{\frac{k}{n+1}+p \frac{1}{n+1}}
{(1-p)+p \frac{k+1}{n+1}}
\le
\frac{\frac{k}{n+1}+(1 -\alpha)\frac{n+1 }{n -k}
 \frac{1}{n+1}}
{\alpha} \nonumber \\
=&
\frac{\frac{k}{n+1}+\frac{1 -\alpha }{n -k}
 }
{\alpha}
=
\frac{k}{ \alpha(n+1)}+\frac{1 -\alpha }{ \alpha(n -k)}.\Label{L1}
\end{align}
Combining \eqref{L2} and \eqref{L1}, we obtain the desired statement.
\end{IEEEproof}

\section{Selftesting of Bell sate}
To discuss the verification of the GHZ state,  
we review the existing result for selftesting of the Bell state by \cite{HH2019}.
To fit our use, we consider the case when the Bell state is given as
$\frac{1}{\sqrt{2}}(|00\rangle_p+|11\rangle_p)$.

We choose a sufficiently large Hilbert spaces
${\cal H}_1''$ and
${\cal H}_2''$ so that
the state on the composite system is the pure state 
$|\psi''\rangle$.
Let 
$\sX''_i$, $\sZ''_i$, $\sA(0)''_i$, $\sA(1)''_i$ be operators on ${\cal H}_i$ for $i=1,2$.

\begin{proposition}\Label{TH17}
When
\begin{align}
&\langle \psi''|\sX''_1 \sX''_2|\psi''\rangle \geq 1-\epsilon, 
 \quad \langle \psi''|-\sZ''_1 \sZ''_2 |\psi''\rangle \geq 1-\epsilon, \Label{e1}\\
&\langle \psi''| \sA(0)''_1(\sX''_2-\sZ''_2)+\sA(1)''_1(\sX''_2+\sZ''_2)
|\psi''\rangle\ge 2\sqrt{2}- \epsilon,\Label{e3}
\end{align}
there exist 
a constant $c_3$ and 
isometries 
$U_1:{\cal H}_1''\to  {\cal H}_1$ and
$U_2:{\cal H}_2''\to  {\cal H}_2$ 
such that the isometry
$U=U_1U_2$ satisfies
\begin{align}
\| U \sX''_1 U^\dagger - \sX_1 \| & \le c_3 \epsilon^{1/2}, ~
\| U \sZ''_1 U^\dagger - \sZ_1 \|  \le c_3 \epsilon^{1/2} \Label{AB1} \\
\| U \sX''_2 U^\dagger - \sX_2 \| & \le c_3 \epsilon^{1/2}, ~
\| U \sZ''_2 U^\dagger - \sZ_2 \|  \le c_3 \epsilon^{1/2} \Label{AB2}.
\end{align}
\end{proposition}

Eq.\eqref{AB1} follows from Proposition 1 of \cite{HH2019}.
While Eq.\eqref{AB2} does not appear in Proposition 1 of \cite{HH2019},
it can be shown by using (E40) and (E41) of Lemma 9.

Now, we apply Proposition \ref{TH17} to the case when 
we prepare $6m+1$ copies of the initial state and 
split them randomly into 6 groups and one final copy.
The procedure is described as follows and is denoted by Protocol \ref{protocol10}:

 \begin{Protocol}                  
\caption{Selftesting of Bell state}    
\Label{protocol10}  
\begin{algorithmic}
\STEPONE
Randomly divide $6m+1$ blocks into 6 groups, in which, the 1st - 6th groups are composed of $m$ blocks.

\STEPTWO
Measure
$\sX'_1$, $\sZ'_1$,
$\sA(0)'_1$, $\sA(0)'_1$, $\sA(1)'_1$, $\sA(1)'_1$ on the system ${\cal H}'_1$
for the 1st - 6th groups.

\STEPTHREE
The corresponding measurements on ${\cal H}'_2$ for the 6 groups are
$\sX'_2$, $\sZ'_2$,
$\sX'_2$, $\sZ'_2$, $\sX'_2$, $\sZ'_2$.
\STEPFOUR
Based on the above measurements, we check the following 3 inequalities for 6 average values:
\begin{align}
&\bA [\sX'_1 \sX'_2] \geq 1-\frac{c_1}{n}, 
\quad \bA [-\sZ'_1 \sZ'_2] \geq 1-\frac{c_1}{n}, \Label{eA1}\\
&\bA [\sA(0)'_1(\sX'_2-\sZ'_2)
+\sA(1)'_1(\sX'_2+\sZ'_2)]\ge \sqrt{2}- \frac{c_1}{\sqrt{n}},\Label{eA3}
\end{align}
Here, the average value in (\ref{eA3}) is calculated from the outcomes 
of the 3rd - 6th groups.
\end{algorithmic}
\end{Protocol}

We apply Conditions \eqref{eA1} and \eqref{eA3} to 
Propositions \ref{TH6} and \ref{TH7}.
Since $\frac{c_2}{n}\le \frac{c_2}{\sqrt{n}}$,
combining Proposition \ref{TH17}, we obtain the following proposition.

\begin{proposition}\Label{TH18}
For significance level $\alpha$ and a constant $c_1$,
there exists a constant $c_4$ to satisfy the following condition.
When the test given in Protocol \ref{protocol10} is passed,
we can guarantee, with significance level $\alpha$, that there exists an isometries 
$U_1:{\cal H}_1''\to  {\cal H}_1$ and
$U_2:{\cal H}_2''\to  {\cal H}_2$ 
such that the isometry
$U=U_1U_2$ satisfies
\begin{align}
\| U \sX''_1 U^\dagger - \sX_1 \| & \le \frac{c_4}{n^{1/4}}, ~
\| U \sZ''_1 U^\dagger - \sZ_1 \|  \le \frac{c_4}{n^{1/4}} \Label{AD1} \\
\| U \sX''_2 U^\dagger - \sX_2 \| & \le \frac{c_4}{n^{1/4}}, ~
\| U \sZ''_2 U^\dagger - \sZ_2 \|  \le \frac{c_4}{n^{1/4}}. \Label{AD2}
\end{align}
\end{proposition}


\section{Extension of Quantum Protocol for Secure Modulo Zero-Sum Randomness
to Case with $\FF_q$}\label{S7B}
Now, we extend our quantum protocol for secure modulo zero-sum randomness
to the case with $\FF_q$.
The following discussion assumes trusted measurement devices. 
Our protocol with untrusted measurement devices cannot be extended to 
the case with $\FF_q$.

When we employ a general finite field $\FF_q$,
the phase basis $\{ |z\rangle_p \}_{z\in\mathbb{F}_q}$ is defined as \cite[Section 8.1.2]{Haya2}
\begin{align*}
|z\rangle_p := \frac{1}{\sqrt{q}} 
\sum_{x\in\mathbb{F}_q} \omega^{-\tr xz} |x\rangle,
\end{align*}
where 
$ |x\rangle $ expresses the computational basis,
$\omega := \exp{\frac{2\pi i}{p}}$ and 
$\tr y$ for $y\in\mathbb{F}_q$ is $\Tr M_y$ where $M_y$ denotes the multiplication map $x \mapsto yx$ with the identificiation of the finite field $\mathbb{F}_q$ with the vector space $\mathbb{F}_p^t$.
Then, the phase GHZ state $
|GHZ\rangle_p:=
\frac{1}{\sqrt{q}}\sum_{z \in \FF_q} |z,\ldots, z \rangle_{p}$
is calculated as
\begin{align}
|GHZ\rangle_p=
\frac{1}{\sqrt{q^{m-1}}} 
\sum_{x_1,\ldots ,x_m\in\mathbb{F}_q:
x_1+\ldots+x_m=0} 
 |x_1,\ldots ,x_m \rangle .
 \end{align}
When all the players apply measurement on the computational basis 
and
the initial state is $|GHZ\rangle_p$,
the sum of $m$ outcomes is zero
and $m-1$ outcomes are subject to the uniform distribution.
Hence, these outcomes satisfy the conditions of secure modulo zero-sum randomness.
That is, when the initial state is guaranteed to be $|GHZ\rangle_p$,
it is guaranteed that the outcomes are secure modulo zero-sum randomness.

When we trust measurement devices, we can employ the following protocol
to verify the state $|GHZ\rangle_p$.

\begin{Protocol}                  
\caption{Verifiable Generation of Secure Modulo Zero-Sum Randomness}    
\label{protocol7B}      
\begin{algorithmic}
\STEPONE[Phase basis check]  
They prepare the system of $2n+1$ copies. 
They randomly choose $n$ copies, and apply the measurement of the phase basis.
If their outcomes are the same, the test is passed.

\STEPTWO[Computational basis check]  
They randomly choose $n$ copies, and apply the measurement of the computational basis.
If the modulo sums of their outcomes are zero, the test is passed.

\STEPTHREE[Generation]  
They apply the measurement of the computational basis to the remaining one copy.
The outcomes are used as secure modulo zero-sum randomness.
 
\end{algorithmic}
\end{Protocol}

\begin{theorem}\label{T2X}
Assume that $\alpha > \frac{1}{2n+1}$ in Protocol \ref{protocol7B}. 
If the test is passed, with significance level $\alpha$, we can guarantee that the
resultant state $\sigma$ on each remaining system satisfies
\begin{align}
\Tr \sigma | GHZ\rangle_p~_p \langle GHZ |
\ge 1 -\frac{1}{\alpha(2n+1)}.\Label{NDR}
\end{align}
\end{theorem}

[Note that the significance level is the maximum passing
probability when malicious Bob sends incorrect states
so that the resultant state $\alpha$ does not satisfy Eq. \eqref{NDR}.] 
The proof of the theorem is given below.
From the theorem and the relation between the fidelity and trace norm [40][(6.106)],
we can conclude the verifiability: if they passed the test,
they can guarantee that
\begin{align}
\|\sigma - | GHZ\rangle_p~_p \langle GHZ |\|_1 \le \frac{1}{\sqrt{\alpha(2n+1)}}
\end{align}
with significance level $\alpha$.
Therefore, when $P_{ideal}$ is the ideal distribution of 
secure modulo zero-sum randomness
and 
$P_{real}$ is the real distribution obtained via the measurement with respect to the computation basis,
we have 
\begin{align}
\|P_{real}-P_{ideal} \|_1 \le \frac{1}{\sqrt{\alpha(2n+1)}}.
\end{align}

\begin{IEEEproof}
We choose a new coordinate $\bar{x}_1, \ldots, \bar{x}_m$
as $\bar{x}_1=x_1+\ldots+x_m$ and $\bar{x}_i=x_i$ for $i=2, \ldots, m$.
We denote the unitary corresponding to this coordinate conversion by $U$.
When a matrix $D$ is applied in the computation basis, 
the conversion on phase basis is given by $(D^{-1})^T$.
Since 
\begin{align}
\left( \left(
\begin{array}{ccccc}
1 &  & & 0 \\
1 & 1 & &  \\
\vdots & \ddots & & \\
1 & 0 & & 1
\end{array}
\right)^{-1} \right)^{T}
=
 \left(
\begin{array}{ccccc}
1 & -1 &\cdots & -1 \\
 & 1 & & 0 \\
 & &\ddots & \\
0&  & & 1
\end{array}
\right),
\end{align}
we have
\begin{align}
U| GHZ\rangle_p=
|0 \rangle |0 , \ldots, 0\rangle_p.
\end{align}
We denote the projection to 
$U^\dagger I\otimes |0,\ldots,0 \rangle_p~_p\langle 0,\ldots,0|  U$ and
$U^\dagger |0 \rangle\langle 0|\otimes I^{\otimes m-1} U$ by 
$\tilde{P}_1$
and $\tilde{P}_2$, respectively.
Then, we find that 
\begin{align}
\tilde{P}_1\tilde{P}_2=
| GHZ\rangle_p~_p \langle GHZ |.
\end{align}
Also, we find that
$\tilde{P}_1$ and $\tilde{P}_2$ are the projections to the subspaces accepting the phase basis check
and the computational basis check, respectively.

We randomly choose one remaining system.
Let $A$ be the random permutation of $\tilde{P}_1^{\otimes n}\otimes \tilde{P}_2^{\otimes n}
\otimes (I-| GHZ\rangle_p~_p \langle GHZ |)$,
which expresses the event that
they accept the test and the state on the remaining system is orthogonal to the state 
$| GHZ\rangle_p~_p \langle GHZ |$.
We define the projection $\bar{P}_i:=\tilde{P}_i-\bar{P}_0$, 
where $\bar{P}_0:=| GHZ\rangle_p~_p \langle GHZ |$ for $i=1,2$.
Also, we define the projection $\bar{P}_3:=
I- | GHZ\rangle_p~_p \langle GHZ |-\bar{P}_1-\bar{P}_2$.
Then, we have 4 orthogonal projections 
$\bar{P}_0,\bar{P}_1,\bar{P}_2,\bar{P}_3$.

Then, we have
\begin{align}
A= \sum_{v \in \{0,1,2,3\}^{2n_1+1}}
\frac{C_1(v)}{C_2(v)}
\bar{P}_{v},
\end{align}
where $\bar{P}_{v}$, $C_1(v)$, and $C_2(v)$ are defined by using the number $N_i(v)$ of $i$ in $v$
as 
\begin{align}
\bar{P}_{v} &:= \bar{P}_{v_1}\otimes \cdots\otimes \bar{P}_{v_{2n_1+1}} \\
C_2(v) &:= {2n_1+1 \choose N_0(v)N_1(v)N_2(v)N_3(v)} \\
C_1(v) &:= \Bigg\{v'
\Bigg| 
\begin{array}{ll}
\bar{P}_{v'} \hbox{ appears in }
\tilde{P}_1^{\otimes n}\otimes \tilde{P}_2^{\otimes n} \otimes (I-| GHZ\rangle_p~_p \langle GHZ |),\\
v' \hbox{ is given as a permutation of }v
\end{array}
\Bigg\}.
\end{align}
Then, we find that 
the maximum eigenvalue of $A$ is $\frac{1}{2n+1}$\footnote{A similar discussion is given \cite[Appendix]{MK}.}.
Since we have $\|A\| \le \frac{1}{2n+1}$,
any initial state $\rho$ satisfies 
$\Tr \rho A \le \frac{1}{2n+1}$.

Now, we assume that the probability accepting the test is less than $\alpha$.
Then, under the condition that they accept the test,
the probability of the event orthogonal to the state $| GHZ\rangle_p~_p \langle GHZ |$
is upper bounded by $\frac{1}{\alpha}\cdot \frac{1}{2n+1}$.
Hence, we obtain the desired statement.
\end{IEEEproof}

\end{document}